\documentstyle[10pt,epsf,amssymb,epsfig,dp_delphititle]{dp_delphi}
\makeindex
\def\DpPaperGroup{EP}
\def\DpPaperRef{98--170}
\def\DpDate{27 October 1998}
\def\DpAuthors{DELPHI Collaboration}
\def\DpSubmit{(Accepted by E. Phys. J. C)}
\def\DpTitle{{Search for lightest neutralino and stau pair production in 
light gravitino scenarios with stau NLSP}}

\begin{document}
%%%%%%%%%%%%%%%%%%%%%%%%%% They are a problem with Coll.Sty ?
\makeatletter
% Collapse citation numbers to ranges.  Non-numeric and undefined labels
% are handled.  No sorting is done.  E.g., 1,3,2,3,4,5,foo,1,2,3,?,4,5
% gives 1,3,2-5,foo,1-3,?,4,5
\newcount\@tempcntc
\def\@citex[#1]#2{\if@filesw\immediate\write\@auxout{\string\citation{#2}}\fi
  \@tempcnta\z@\@tempcntb\m@ne\def\@citea{}\@cite{\@for\@citeb:=#2\do
    {\@ifundefined
       {b@\@citeb}{\@citeo\@tempcntb\m@ne\@citea\def\@citea{,}{\bf ?}\@warning
       {Citation `\@citeb' on page \thepage \space undefined}}%
    {\setbox\z@\hbox{\global\@tempcntc0\csname b@\@citeb\endcsname\relax}%
     \ifnum\@tempcntc=\z@ \@citeo\@tempcntb\m@ne
       \@citea\def\@citea{,}\hbox{\csname b@\@citeb\endcsname}%
     \else
      \advance\@tempcntb\@ne
      \ifnum\@tempcntb=\@tempcntc
      \else\advance\@tempcntb\m@ne\@citeo
      \@tempcnta\@tempcntc\@tempcntb\@tempcntc\fi\fi}}\@citeo}{#1}}
\def\@citeo{\ifnum\@tempcnta>\@tempcntb\else\@citea\def\@citea{,}%
  \ifnum\@tempcnta=\@tempcntb\the\@tempcnta\else
   {\advance\@tempcnta\@ne\ifnum\@tempcnta=\@tempcntb \else \def\@citea{--}\fi
    \advance\@tempcnta\m@ne\the\@tempcnta\@citea\the\@tempcntb}\fi\fi}
 
\makeatother
%%%%%%%%%%%%%%%%%%%%%%%%%% ??????????????????????????????????
% Generate the title page
\begin{titlepage}
\pagenumbering{roman}
\CERNpreprint{\DpPaperGroup}{\DpPaperRef} % Reference of the paper
\date{{\small\DpDate}} % Date of the paper
\title{\DpTitle} % Title of the paper
\address{\DpAuthors} % General name of the author(s)
\begin{shortabs} % Start the abstract
\noindent
%   abstract.tex
%
\newcommand{\GeV}     {\mbox{${\mathrm{GeV}}$}}
\newcommand{\GeVcc}   {\mbox{${\mathrm{GeV}}/{\mathrm{c}}^2$}}
\newcommand{\eVcc}    {\mbox{${\mathrm{eV}}/{\mathrm{c}}^2$}}

\noindent

Promptly decaying lightest neutralinos and long-lived staus
are searched for in the context of light gravitino scenarios.
It is assumed that the stau is the next to lightest supersymmetric 
particle (NLSP) and that the lightest neutralino is the next to NLSP (NNLSP).
Data collected with the Delphi
detector at centre-of-mass energies from 161 to 183~\GeV\ are analysed.
No evidence of the production of these particles is found.
Hence, lower mass limits for both kinds of particles are set at 95\%
C.L.. The mass of gaugino-like neutralinos is found to be greater than 
71.5 GeV/$c^2$.
In the search for long-lived stau, masses
less than 70.0 to 77.5~\GeVcc\ are excluded
for gravitino masses from 10 to 150~\eVcc . Combining this search
with the searches for stable heavy leptons and Minimal Supersymmetric
Standard Model
staus a lower limit of 68.5~\GeVcc\  may be set for the stau mass 
independent of the mass of the gravitino.
\end{shortabs}
\vfill
\begin{center}
\DpSubmit \ % Horrible hack to allow to have DpSubmit empty
%\DpComment \ \\
%\DpEMail \ \\
\end{center}
\vfill
\clearpage
\headsep 10.0pt
\addtolength{\textheight}{10mm}
\addtolength{\footskip}{-5mm}
\begingroup
% Commands to process the author names
%
\newcommand{\DpName}[2]{\hbox{#1$^{\ref{#2}}$},\hfill}
\newcommand{\DpNameTwo}[3]{\hbox{#1$^{\ref{#2},\ref{#3}}$},\hfill}
\newcommand{\DpNameThree}[4]{\hbox{#1$^{\ref{#2},\ref{#3},\ref{#4}}$},\hfill}
\newskip\Bigfill \Bigfill = 0pt plus 1000fill
\newcommand{\DpNameLast}[2]{\hbox{#1$^{\ref{#2}}$}\hspace{\Bigfill}}
%
%CD\small
\footnotesize
\noindent
\DpName{P.Abreu}{LIP}
\DpName{W.Adam}{VIENNA}
\DpName{T.Adye}{RAL}
\DpName{P.Adzic}{DEMOKRITOS}
\DpName{T.Aldeweireld}{AIM}
\DpName{G.D.Alekseev}{JINR}
\DpName{R.Alemany}{VALENCIA}
\DpName{T.Allmendinger}{KARLSRUHE}
\DpName{P.P.Allport}{LIVERPOOL}
\DpName{S.Almehed}{LUND}
% ...
\normalsize
\endgroup
\titlefoot{Department of Physics and Astronomy, Iowa State
     University, Ames IA 50011-3160, USA
    \label{AMES}}
\titlefoot{Physics Department, Univ. Instelling Antwerpen,
     Universiteitsplein 1, BE-2610 Wilrijk, Belgium \\
     \indent~~and IIHE, ULB-VUB,
     Pleinlaan 2, BE-1050 Brussels, Belgium \\
     \indent~~and Facult\'e des Sciences,
     Univ. de l'Etat Mons, Av. Maistriau 19, BE-7000 Mons, Belgium
    \label{AIM}}
\titlefoot{Physics Laboratory, University of Athens, Solonos Str.
     104, GR-10680 Athens, Greece
    \label{ATHENS}}
\titlefoot{Department of Physics, University of Bergen,
     All\'egaten 55, NO-5007 Bergen, Norway
    \label{BERGEN}}
\titlefoot{Dipartimento di Fisica, Universit\`a di Bologna and INFN,
     Via Irnerio 46, IT-40126 Bologna, Italy
    \label{BOLOGNA}}
\titlefoot{Centro Brasileiro de Pesquisas F\'{\i}sicas, rua Xavier Sigaud 150,
     BR-22290 Rio de Janeiro, Brazil \\
     \indent~~and Depto. de F\'{\i}sica, Pont. Univ. Cat\'olica,
     C.P. 38071 BR-22453 Rio de Janeiro, Brazil \\
     \indent~~and Inst. de F\'{\i}sica, Univ. Estadual do Rio de Janeiro,
     rua S\~{a}o Francisco Xavier 524, Rio de Janeiro, Brazil
    \label{BRASIL}}
\titlefoot{Comenius University, Faculty of Mathematics and Physics,
     Mlynska Dolina, SK-84215 Bratislava, Slovakia
    \label{BRATISLAVA}}
\titlefoot{Coll\`ege de France, Lab. de Physique Corpusculaire, IN2P3-CNRS,
     FR-75231 Paris Cedex 05, France
    \label{CDF}}
\titlefoot{CERN, CH-1211 Geneva 23, Switzerland
    \label{CERN}}
\titlefoot{Institut de Recherches Subatomiques, IN2P3 - CNRS/ULP - BP20,
     FR-67037 Strasbourg Cedex, France
    \label{CRN}}
\titlefoot{Institute of Nuclear Physics, N.C.S.R. Demokritos,
     P.O. Box 60228, GR-15310 Athens, Greece
    \label{DEMOKRITOS}}
\titlefoot{FZU, Inst. of Phys. of the C.A.S. High Energy Physics Division,
     Na Slovance 2, CZ-180 40, Praha 8, Czech Republic
    \label{FZU}}
\titlefoot{Dipartimento di Fisica, Universit\`a di Genova and INFN,
     Via Dodecaneso 33, IT-16146 Genova, Italy
    \label{GENOVA}}
\titlefoot{Institut des Sciences Nucl\'eaires, IN2P3-CNRS, Universit\'e
     de Grenoble 1, FR-38026 Grenoble Cedex, France
    \label{GRENOBLE}}
\titlefoot{Helsinki Institute of Physics, HIP,
     P.O. Box 9, FI-00014 Helsinki, Finland
    \label{HELSINKI}}
\titlefoot{Joint Institute for Nuclear Research, Dubna, Head Post
     Office, P.O. Box 79, RU-101 000 Moscow, Russian Federation
    \label{JINR}}
\titlefoot{Institut f\"ur Experimentelle Kernphysik,
     Universit\"at Karlsruhe, Postfach 6980, DE-76128 Karlsruhe,
     Germany
    \label{KARLSRUHE}}
\titlefoot{Institute of Nuclear Physics and University of Mining and Metalurgy,
     Ul. Kawiory 26a, PL-30055 Krakow, Poland
    \label{KRAKOW}}
\titlefoot{Universit\'e de Paris-Sud, Lab. de l'Acc\'el\'erateur
     Lin\'eaire, IN2P3-CNRS, B\^{a}t. 200, FR-91405 Orsay Cedex, France
    \label{LAL}}
\titlefoot{School of Physics and Chemistry, University of Lancaster,
     Lancaster LA1 4YB, UK
    \label{LANCASTER}}
\titlefoot{LIP, IST, FCUL - Av. Elias Garcia, 14-$1^{o}$,
     PT-1000 Lisboa Codex, Portugal
    \label{LIP}}
\titlefoot{Department of Physics, University of Liverpool, P.O.
     Box 147, Liverpool L69 3BX, UK
    \label{LIVERPOOL}}
\titlefoot{LPNHE, IN2P3-CNRS, Univ.~Paris VI et VII, Tour 33 (RdC),
     4 place Jussieu, FR-75252 Paris Cedex 05, France
    \label{LPNHE}}
\titlefoot{Department of Physics, University of Lund,
     S\"olvegatan 14, SE-223 63 Lund, Sweden
    \label{LUND}}
\titlefoot{Universit\'e Claude Bernard de Lyon, IPNL, IN2P3-CNRS,
     FR-69622 Villeurbanne Cedex, France
    \label{LYON}}
\titlefoot{Univ. d'Aix - Marseille II - CPP, IN2P3-CNRS,
     FR-13288 Marseille Cedex 09, France
    \label{MARSEILLE}}
\titlefoot{Dipartimento di Fisica, Universit\`a di Milano and INFN,
     Via Celoria 16, IT-20133 Milan, Italy
    \label{MILANO}}
\titlefoot{Niels Bohr Institute, Blegdamsvej 17,
     DK-2100 Copenhagen {\O}, Denmark
    \label{NBI}}
\titlefoot{NC, Nuclear Centre of MFF, Charles University, Areal MFF,
     V Holesovickach 2, CZ-180 00, Praha 8, Czech Republic
    \label{NC}}
\titlefoot{NIKHEF, Postbus 41882, NL-1009 DB
     Amsterdam, The Netherlands
    \label{NIKHEF}}
\titlefoot{National Technical University, Physics Department,
     Zografou Campus, GR-15773 Athens, Greece
    \label{NTU-ATHENS}}
\titlefoot{Physics Department, University of Oslo, Blindern,
     NO-1000 Oslo 3, Norway
    \label{OSLO}}
\titlefoot{Dpto. Fisica, Univ. Oviedo, Avda. Calvo Sotelo
     s/n, ES-33007 Oviedo, Spain
    \label{OVIEDO}}
\titlefoot{Department of Physics, University of Oxford,
     Keble Road, Oxford OX1 3RH, UK
    \label{OXFORD}}
\titlefoot{Dipartimento di Fisica, Universit\`a di Padova and
     INFN, Via Marzolo 8, IT-35131 Padua, Italy
    \label{PADOVA}}
\titlefoot{Rutherford Appleton Laboratory, Chilton, Didcot
     OX11 OQX, UK
    \label{RAL}}
\titlefoot{Dipartimento di Fisica, Universit\`a di Roma II and
     INFN, Tor Vergata, IT-00173 Rome, Italy
    \label{ROMA2}}
\titlefoot{Dipartimento di Fisica, Universit\`a di Roma III and
     INFN, Via della Vasca Navale 84, IT-00146 Rome, Italy
    \label{ROMA3}}
\titlefoot{DAPNIA/Service de Physique des Particules,
     CEA-Saclay, FR-91191 Gif-sur-Yvette Cedex, France
    \label{SACLAY}}
\titlefoot{Instituto de Fisica de Cantabria (CSIC-UC), Avda.
     los Castros s/n, ES-39006 Santander, Spain
    \label{SANTANDER}}
\titlefoot{Dipartimento di Fisica, Universit\`a degli Studi di Roma
     La Sapienza, Piazzale Aldo Moro 2, IT-00185 Rome, Italy
    \label{SAPIENZA}}
\titlefoot{Inst. for High Energy Physics, Serpukov
     P.O. Box 35, Protvino, (Moscow Region), Russian Federation
    \label{SERPUKHOV}}
\titlefoot{J. Stefan Institute, Jamova 39, SI-1000 Ljubljana, Slovenia
     and Department of Astroparticle Physics, School of\\
     \indent~~Environmental Sciences, Kostanjeviska 16a, Nova Gorica,
     SI-5000 Slovenia, \\
     \indent~~and Department of Physics, University of Ljubljana,
     SI-1000 Ljubljana, Slovenia
    \label{SLOVENIJA}}
\titlefoot{Fysikum, Stockholm University,
     Box 6730, SE-113 85 Stockholm, Sweden
    \label{STOCKHOLM}}
\titlefoot{Dipartimento di Fisica Sperimentale, Universit\`a di
     Torino and INFN, Via P. Giuria 1, IT-10125 Turin, Italy
    \label{TORINO}}
\titlefoot{Dipartimento di Fisica, Universit\`a di Trieste and
     INFN, Via A. Valerio 2, IT-34127 Trieste, Italy \\
     \indent~~and Istituto di Fisica, Universit\`a di Udine,
     IT-33100 Udine, Italy
    \label{TU}}
\titlefoot{Univ. Federal do Rio de Janeiro, C.P. 68528
     Cidade Univ., Ilha do Fund\~ao
     BR-21945-970 Rio de Janeiro, Brazil
    \label{UFRJ}}
\titlefoot{Department of Radiation Sciences, University of
     Uppsala, P.O. Box 535, SE-751 21 Uppsala, Sweden
    \label{UPPSALA}}
\titlefoot{IFIC, Valencia-CSIC, and D.F.A.M.N., U. de Valencia,
     Avda. Dr. Moliner 50, ES-46100 Burjassot (Valencia), Spain
    \label{VALENCIA}}
\titlefoot{Institut f\"ur Hochenergiephysik, \"Osterr. Akad.
     d. Wissensch., Nikolsdorfergasse 18, AT-1050 Vienna, Austria
    \label{VIENNA}}
\titlefoot{Inst. Nuclear Studies and University of Warsaw, Ul.
     Hoza 69, PL-00681 Warsaw, Poland
    \label{WARSZAWA}}
\titlefoot{Fachbereich Physik, University of Wuppertal, Postfach
     100 127, DE-42097 Wuppertal, Germany
    \label{WUPPERTAL}}
\titlefoot{On leave of absence from IHEP Serpukhov
    \label{MILAN-SERPOU}}
\titlefoot{Now at University of Florida
    \label{FLORIDA}}
\addtolength{\textheight}{-10mm}
\addtolength{\footskip}{5mm}
\clearpage
\headsep 30.0pt
\end{titlepage}
%%%%%%%%%%%%%%%%%%%%%%%%%
%
% Change for the document body
%%%CD\pagestyle{heading} % for page numbering
\pagenumbering{arabic} % page numbering in number
\setcounter{footnote}{0} %
\large
%   document.tex
%
%%% put your own definitions here:
\newcommand{\stau}   {$\tilde{\tau}$}
\newcommand{\stuno}  {$\tilde{\tau}_1$}
\newcommand{\nuno}   {$\tilde{\chi}^0_1$}
\newcommand{\ra}     {\rightarrow}
\newcommand{\eeto}    {\mbox{$ {\mathrm e}^+ {\mathrm e}^-\! \ra\ $}}
\newcommand{\ee}      {\mbox{$ {\mathrm e}^+ {\mathrm e}^-$}}
\newcommand{\qqbar}  {$q\bar{q}$}
\newcommand{\Wp}     {\mbox{$ {\mathrm W}^+$}}
\newcommand{\Wm}     {\mbox{$ {\mathrm W}^-$}}
\newcommand{\Zn}      {\mbox{$ {\mathrm Z}$}}
\newcommand{\Wev}     {\mbox{$ {\mathrm{W e}} \nu_{\mathrm e}$}}
\newcommand{\Zvv}     {\mbox{$ \Zn \nu \bar{\nu}$}}
\newcommand{\Zee}     {\mbox{$ \Zn \ee$}}
\newcommand{\MeV}     {\mbox{$ {\mathrm{MeV}}                             $}}
\newcommand{\MeVc}    {\mbox{$ {\mathrm{MeV}}/c                           $}}
\newcommand{\MeVcc}   {\mbox{$ {\mathrm{MeV}}/c^2                         $}}
\newcommand{\GeV}     {\mbox{$ {\mathrm{GeV}}                             $}}
\newcommand{\GeVc}    {\mbox{$ {\mathrm{GeV}}/c                           $}}
\newcommand{\GeVcc}   {\mbox{$ {\mathrm{GeV}}/c^2                         $}}
\newcommand{\eVcc}    {\mbox{$ {\mathrm{eV}}/c^2                          $}}
\newcommand{\TeV}     {\mbox{$ {\mathrm{TeV}}                             $}}
\newcommand{\etal}  {\mbox{\it et al.}}
\def\NPB#1#2#3{{\rm Nucl.~Phys.} {\bf{B#1}} (19#2) #3}
\def\PLB#1#2#3{{\rm Phys.~Lett.} {\bf{B#1}} (19#2) #3}
\def\PRD#1#2#3{{\rm Phys.~Rev.} {\bf{D#1}} (19#2) #3}
\def\PRL#1#2#3{{\rm Phys.~Rev.~Lett.} {\bf{#1}} (19#2) #3}
\def\ZPC#1#2#3{{\rm Z.~Phys.} {\bf C#1} (19#2) #3}
\def\PTP#1#2#3{{\rm Prog.~Theor.~Phys.} {\bf#1}  (19#2) #3}
\def\MPL#1#2#3{{\rm Mod.~Phys.~Lett.} {\bf#1} (19#2) #3}
\def\PR#1#2#3{{\rm Phys.~Rep.} {\bf#1} (19#2) #3}
\def\RMP#1#2#3{{\rm Rev.~Mod.~Phys.} {\bf#1} (19#2) #3}
\def\HPA#1#2#3{{\rm Helv.~Phys.~Acta} {\bf#1} (19#2) #3}
\def\NIMA#1#2#3{{\rm Nucl.~Instr.~and~Meth.} {\bf#1} (19#2) #3}
\def\CPC#1#2#3{{\rm Comp.~Phys.~Comm.} {\bf#1} (19#2) #3}

\section{Introduction}
\label{sec:intro}
%
%If supersymmetry at the electroweak scale is established, one of the most
%important questions to address experimentally is the scale and mechanism of
%supersymmetry breaking. 
In models including supersymmetry (SUSY),
it is often assumed that the messengers of
supersymmetry breaking couple to
the observable sector with interactions of gravitational 
strength and that the SUSY
breaking scale in the hidden sector is of the order of
$10^{11}$~\GeV. An alternative
possibility is that supersymmetry is broken at some lower scale 
(below $10^{7}$~\GeV), and that the ordinary gauge interaction
acts as the messenger of supersymmetry breaking \cite{Dine1,Dine2}.
%In local supersymmetry, the Goldstino becomes the
%longitudinal component
%of the gravitino ($\tilde{G}$).
In this case, the gravitino, $\tilde{G}$, is naturally the lightest 
supersymmetric particle
(LSP) and the lightest Standard Model superpartner is the next to lightest
supersymmetric particle (NLSP).
Thus, the NLSP is unstable and decays to its
Standard Model (SM) partner and a gravitino.

Since the gravitino couplings are in general, 
with the exception of the so-called
ultra-light gravitino scenarios,
suppressed compared to electroweak and
strong interactions, decays to the gravitino are 
in general only relevant for the NLSP
and therefore the production and decay of supersymmetric
particles at high energy colliders would ge\-ne\-ra\-lly 
take place through Standard
Model
couplings~\footnote{One exception to this rule being the process
$e^+e^-\to Z^*/{\gamma}^{*}\to \tilde{G}\tilde{\chi}^0_1$
for the case of ultra-light $\tilde{G}$ scenarios.}.
%or possibly, the decay of the lightest neutralino into a photon and a 
%gravitino instead of into
%a stau and a tau
%for the case of ultra-light $\tilde{G}$ scenarios.}.
The supersymmetric particles decay into the NLSP, which
eventually decays to its SM partner and a gravitino. The specific
signatures of such decays depend crucially on the quantum numbers and
composition of the NLSP. 
%This opens up the possibility to detect
%$\tilde{\chi}^0_1 \tilde{\chi}^0_1$\ pair production from its
%decay products, unlike in SUGRA models.

Although most of the attention has been focused on the case where the
neutralino is the NLSP, it is also possible that the NLSP is any
other sparticle, and in particular
a charged
slepton. The number of %$5+\bar{5}$ 
generations of supersymmetry breaking 
messengers
in minimal models, $n$, 
determines over most of the parameter space which particle is
the NLSP~\cite{Bagger,Dutta,francesca,giudice}. For
example, for one generation of messengers, %$5+\bar{5}$ 
the
lightest neutralino tends to be the
NLSP,
while for two or more  generations, right handed sleptons are favoured.
Moreover, when left-right sfermion mixing~\cite{bartl} occurs, 
the corresponding
$\tilde{\tau}$\ state, \stuno , becomes the NLSP.

Throughout this work, it is assumed that the \stuno\ is the NLSP and
that the lightest neutralino, \nuno , is the {\it next-to}-NLSP (NNLSP).
The \stuno\ width is given (independently of the $\tilde{\tau}$\ 
mixing) by the
two-body equation:
\begin{equation}
\label{gamma}
\Gamma (\tilde{\tau}_1 \rightarrow \tau + \tilde{G} )
= \frac{m^5_{\tilde{\tau}_1}}{48 \pi M^2_p m^2_{\tilde{G}}}
\end{equation}
where $m_{\tilde{\tau}_1}$ is the mass of the $\tilde{\tau}_1$, 
$m_{\tilde{G}}$ is the mass of the $\tilde{G}$~and $M_p$ is the Planck mass
($2.4 \times 10^{18}$ \GeV). In the last equation, the 
mass of the $\tau$\ has been neglected.
The mean decay length obtained from equation (\ref{gamma}):
\begin{equation}
L = 1.76 \times 10^{-3} (E^2/m^2_{\tilde{\tau}_1}-1)^{\frac{1}{2}}
\left ( \frac{m_{\tilde{\tau}_1}}{100 \, {\rm GeV/c}^2} \right )^{-5}
\left ( \frac{m_{\tilde{G}}}{1\, \rm eV/c^2}\right )^{2} \;\; {\mathrm cm,}
\end{equation}
depends strongly on $m_{\tilde{\tau}}$,  $m_{\tilde{G}}$ and the energy of the
 \stuno, $E$.  
The dependence of the mean decay length, $L$, on $m_{\tilde{G}}$
could be also interpreted in terms of the supersymmetry breaking scale,
$\sqrt{F}$, through the relation:
\begin{equation}
m_{\tilde{G}} = \frac{F}{\sqrt{3}M_p} \simeq 2.5
\left ( \frac{\sqrt{F}}{100 \,\rm TeV}\right )^2 \;\;{\rm eV/c}^2
\end{equation}
For $\sqrt{F} \lesssim 1000$~\TeV\ 
($m_{\tilde{G}}\lesssim  250$ eV), the decay can take place
within the detector. This range of $\sqrt{F}$ is in fact consistent with
astrophysical and cosmological considerations \cite{Dinopoulos0}.

Two searches are presented here. Firstly, \nuno~pair production with 
$\tilde{\chi}^0_1$ decaying to $\tilde{\tau}_1 \tau$ and 
$\tilde{\tau}_1$ then 
 decaying promptly into
$\tau \tilde{G}$.
%Long lived decays of the \stau\ will be
%studied in a future work. The search for direct pair production of {\stau}'s
%is described elsewhere~\cite{ref:flying}. The former search is complementary to
%the later in regions where \stau - pair production has small cross-section.
The signature of the signal would be
four $\tau$'s with missing energy
and momentum from the two gravitinos (in addition to the energy and momentum
carried away by the neutrinos of the decay of the $\tau$'s).

The second search concerns $\tilde{\tau}_1$\ pair production followed
by the decays $\tilde{\tau}_1 \rightarrow \tau \tilde{G}$
within the detector volume.
% assuming 
%conservatively  the $\tilde{\tau}_1$ to be right-handed.
%The stau, however, could be degenerate with other sleptons which would
%then contribute to this search. This analysis therefore considers the most
%conservative hypothesis.
The signature of such an event will be a
track of a charged particle
with a kink or a decay vertex when the
$\tilde{\tau}_1$ decays inside the tracking devices. If the decay length is
too short (small $m_{\tilde{G}}$) to allow for the reconstruction of the
$\tilde{\tau}_1$ track, only the decay products of the $\tau$ will be seen in
the detector, and the search will then be based on track impact parameter.
However, if the decay takes place outside the tracking devices
(large $m_{\tilde{G}}$),
the signature will be that of a heavy charged particle already studied
in DELPHI~\cite{Heavyparticles}. For very light $m_{\tilde{G}}$ the limits
from the search for MSSM stau can be applied~\cite{stau-pair}.
All these searches have been combined to obtain a limit on
$m_{\tilde{\tau}_R}$ independent of the
$\tilde{G}$ mass.

The data samples and event selections are respectively described in 
sections~\ref{experimentalprocedure} 
and 3, while the results are presented in
section~\ref{sec:resultados}. 
It will be seen in section~\ref{sec:resultados} that these two searches,  
together with those for 
$\tilde{\chi}^0_1\to \gamma \tilde{G}$~\cite{2gamma} 
(in the $\tilde{\chi}^0_1$ NLSP scenario)
and promptly decaying \stuno\ pair production~\cite{stau-pair} 
complement each other for different
domains of the gravitino mass.

%
%------------------------------------------------------------
% EVENT SELECTION
%------------------------------------------------------------
%
\section{Event sample and experimental procedure}
\label{experimentalprocedure}

The search for neutralino pair production is based 
on data collected by the DELPHI experiment
 during 1996 and 1997 at
centre-of-mass energies of 161, 172 and 183~\GeV. 
The total integrated luminosities for the three centre-of-mass energies 
are 9.7, 10.4 and 53.9 ${\mathrm pb}^{-1}$\ respectively.  
The search for stau pair production with
big impact parameters and secondary vertices
is based on data collected by the DELPHI experiment
 during 1997 since the results obtained with the 
data collected in 1995 (at $\sqrt{s} = 130-136$~\GeV)
and 1996 are already published in~\cite{ref:flying}. The present analysis 
updates those results.
The search for stau pair production with
small impact parameters is based on data collected  
from 1995 to 1997.
A detailed
description of the DELPHI detector can be found in \cite{detector} and its
performance in \cite{performance}.

To evaluate the signal efficiencies and background contaminations,
events were ge\-ne\-ra\-ted using different programs, all
relying on {\tt JETSET} 7.4 \cite{JETSET}, tuned
to LEP~1 data \cite{TUNE} for quark fragmentation.
The program {\tt SUSYGEN} \cite{SUSYGEN} was used to generate
the neutralino pair events and their subsequent decay products. 
In order to compute detection efficiencies, a
total of 3000, 10000 and 14000 events were generated 
with centre-of-mass energies of 161, 172 and 183~\GeV\ respectively, 
and masses 
47~\GeVcc$\leq m_{\tilde{\tau}_1}+2$~\GeVcc~$\leq m_{\tilde{\chi}^0_1} \leq 
  \sqrt{s}/2$.
A stau pair sample of 18000 events 
(subdivided in 15 samples) was produced 
with {\tt PYTHIA 5.7} \cite{JETSET} at 183~\GeV~centre-of-mass energy,
the staus having mean decay lengths from  0.25 to 1000~cm and masses
from 40 to 90~\GeVcc.
Another sample of 35000 stau pairs was produced with {\tt SUSYGEN} for the 
small impact parameter search (see below), with centre-of-mass energies 
ranging from 130~\GeV\ up to 183~\GeV.

The background process \eeto\qqbar ($n\gamma$) was generated with
{\tt PYTHIA 5.7}, while {\tt DYMU3} \cite{DYMU3} and
{\tt KORALZ} \cite{KORALZ} were used
for $\mu^+\mu^-(\gamma)$ and $\tau^+\tau^-(\gamma)$,
respectively.
The generator
of reference~\cite{BAFO} was used for \eeto\ee\ events.

Processes leading to four-fermion final states,
$(\Zn/\gamma)^*(\Zn/\gamma)^*$, where * means of-the-mass-shell, 
$\Wp \Wm $, \Wev\ and \Zee,
were also generated using {\tt PYTHIA}.
%The cut on the invariant mass
%of the virtual $(\Zn/\gamma)^*$ in  the $(\Zn/\gamma)^*(\Zn/\gamma)^*$
%process
%was set at  2~\GeVcc, in order
%to determine the background from low mass \ffbar\ pairs.
The calculation of the four-fermion
background was verified using the program {\tt EXCALIBUR}
\cite{EXCALIBUR}, which consistently
takes into account all amplitudes leading to a given four-fermion
final state. 

Two-photon interactions leading to hadronic final states
%leading to hadronic and leptonic final states
were generated using {\tt TWOGAM} \cite{TWOGAM}, separating the VDM, QPM and
QCD components.
The generators
of Berends, Daverveldt and Kleiss \cite{BDK} were used for the leptonic
final states.

The cosmic radiation background was studied using the data collected
before the beginning of the 1997 LEP run.

The generated signal and background events were passed through the
detailed simulation~\cite{performance}
%~\cite{delsim} 
of the DELPHI detector 
%\cite{detector} 
and then processed
with the same reconstruction and analysis programs used for real 
data.
%The number of simulated events from different background processes
%corresponds to
%several times
%that of the real data.

\section{Data selection}
\label{dataselection}
\subsection{Neutralino pair production}
\label{neutra}

In this section, the selections used to search for the process
$e^+e^- \to \tilde{\chi}^0_1 \tilde{\chi}^0_1 \to 
\tilde{\tau}_1 \tau \tilde{\tau}_1 \tau  \to \tau \tilde{G} \tau \tau 
\tilde{G} \tau $\ are presented. 

%The preselection of tracks of charged particles
% and neutral clusters is described 
%in reference~\cite{chargi}.
The reconstructed tracks of charged particles were 
required to have momenta above 100~\MeVc\ and impact parameters below 4~cm
in the transverse plane and below 10~cm in the lon\-gi\-tu\-di\-nal
direction. The relative error on the measurement of the momentum was to be
smaller than 100\%.
Clusters in the calorimeters were interpreted as neutral
particles if they were not associated to charged particles and if their
energy exceeded 100~\MeV. 
All charged and neutral particles that satisfy these criteria are 
considered good particles and they are used to compute the relevant event 
quantities. 
To assure good quality of the data, the ratio of good to  
total number of tracks was required to be above 
0.7. Tracks that did not pass quality selection but had an
associated calorimetric energy of at least 2~\GeV~had their angles 
taken from those of the track, but their momentum
was recomputed as that of the calorimetric measurement. 
Such tracks were
not included in the good sample. Events had to have between 
four and ten good charged particle tracks. 
%Our definition of 
%a good track does not include recovered tracks, but they are used in 
%the calculation of all kinematic variables.
In addition, it was required that the thrust be less than 0.99. 
The transverse momentum, computed as the transverse component with respect 
  to the beam axis of the vector sum of the momenta of good charged and 
neutral particles, $p_T$, had to
be bigger than 3~\GeVc. And the absolute value of the cosine of the polar 
angle of the missing momentum vector be less than 0.95. 
Very forward-going events
% and part of the $\gamma - \gamma$\ background
were eliminated by requiring
that the energy in a cone of $30^{\circ}$, $E_{30}$, around 
the beam-pipe to be less than 
70\% of the total visible energy, $E_{vis}$.
With this preselection, the total number of simulated background events and real data events 
was reduced by a factor 
of about 6000. Only events passing these pre-selections were 
analysed further.

\begin{figure}[htbp]\centering
\epsfxsize=16.0cm
\centerline{\epsffile{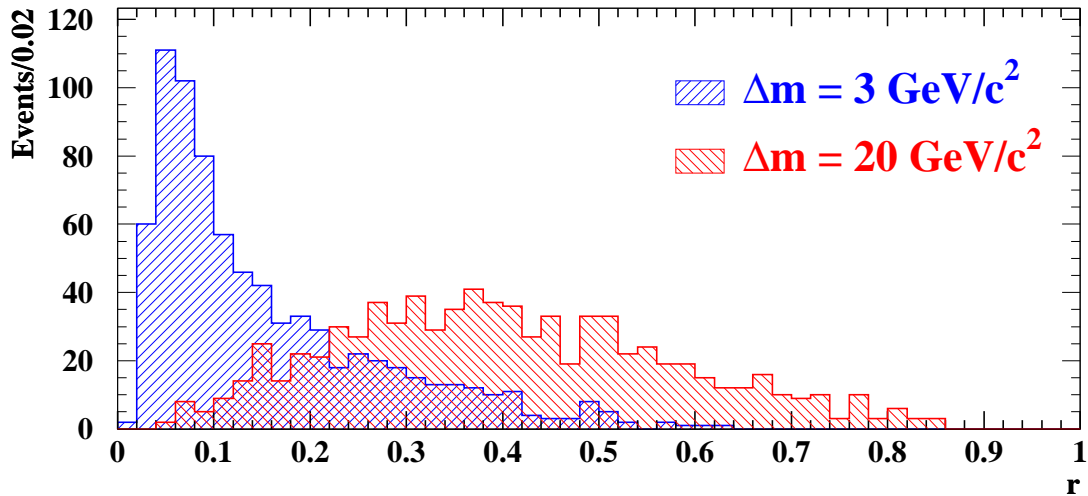}}
\vspace{-8cm}
\caption{Two examples of the
distribution of the variable $r$ (see text). 
The positive-slope hatched histogram shows $r$\ for   
$\Delta m = 3$~\GeVcc. 
The negative-slope hatched histogram shows $r$\ for   
$\Delta m = 20$~\GeVcc.}
\label{fig-ratio}
\end{figure}

The selection takes advantage of the fact that 
signal events can be separated into two different kinematic regions of
the ($m_{\tilde{\chi}^0_1}$,$m_{\tilde{\tau}_1}$) space: when the 
mass difference $\Delta m =   m_{\tilde{\chi}^0_1} - m_{\tilde{\tau}_1}$\ 
is bigger than about 10~\GeVcc, all four $\tau$'s carry similar momenta. 
When the difference becomes smaller, the two  $\tau$'s coming from the 
decay of the $\tilde{\tau}_1$\ tend to be the most energetic, 
increasingly so as the ${\tilde{\chi}^0_1}$\ mass increases. The Durham 
algorithm~\cite{Durham} was used to 
divide the event in four jets 
by allowing $y_{cut}$ to vary as a free variable. 
Numbering the jets from 1 to 4 with 
${\rm E}_1 > {\rm E}_2 > {\rm E}_3 > {\rm E}_4$, a variable
$r$ was defined as:
\begin{equation}
r = \frac{{\rm E}_3 + {\rm E}_4}{{\rm E}_1 + {\rm E}_2} \ .
\end{equation}
An example of the distribution of $r$\ for simulated samples with 
two values of $\Delta m$ can be seen 
in figure~\ref{fig-ratio}. It should be noticed that the distribution of $r$\
shifts towards lower values with increasing neutralino masses. 
The simulated background samples were then divided into 
two samples above and below $r = 0.1$ and different requirements were
imposed in the two cases.

\vspace{1cm}
Two sets of cuts were applied in order to reduce the 
$\gamma\gamma$\ and ${\rm f}{\bar{{\rm f}}}(\gamma)$\ backgrounds
and a third set of cuts to select events according to their topology:

\begin{itemize}
\item[1-] {\underline{Cuts against $\gamma\gamma$\ backgrounds}:
the transverse energy, ${\rm E_T}$, should be bigger than
11~\GeV\ for $r>0.1$\ (${\rm E_T} >$ 12~\GeV\ for $r\leq0.1$).
The energy in a cone of $30^{\circ}$
around the beam axis was further 
restricted to be less than 
60\% of the total visible energy to avoid possible bias from the
Monte Carlo samples. The missing mass should be smaller 
than 0.88$\sqrt{s}$\ (0.9$\sqrt{s}$).
The momentum of the charged particle with largest momentum 
should 
be bigger than 4~\GeVc\ (3~\GeVc).
These cuts reduced the $\gamma\gamma$ background  
by a factor of the order of 30.}

\item[2-] {\underline{Cuts against ${\rm f}{\bar{{\rm f}}}(\gamma)$\  
backgrounds}:
the number of good tracks should be smaller than 7 (9).
The maximum thrust was further reduced from 0.99 to 0.975. 
Dividing each event into two jets with the Durham algorithm, 
its
acoplanarity should be bigger than $8^{\circ}$. The missing mass of the 
events should be bigger 
than 0.3$\sqrt{s}$.
After these cuts, the ${\rm f}{\bar{{\rm f}}}(\gamma)$\  
background was reduced by a 
factor of the order of 15.}

\item[3-] {\underline{Cuts based on topology}:
signal events tend naturally to cluster into a 4-jet topology.  
All jets should 
be at least $17^{\circ}$\ away from the beam direction. 
When reduced by the jet algorithm into a 2-jet configuration, 
the charged particles belonging to each of these jets should be 
in a cone broader than $20^{\circ}$. Finally, the axes of
each of the four jets should be separated from the others at 
least by $8^{\circ}$\ ($4^{\circ}$).}
\end{itemize}

\begin{figure}[htbp]\centering
\epsfxsize=16.0cm
\centerline{\epsffile{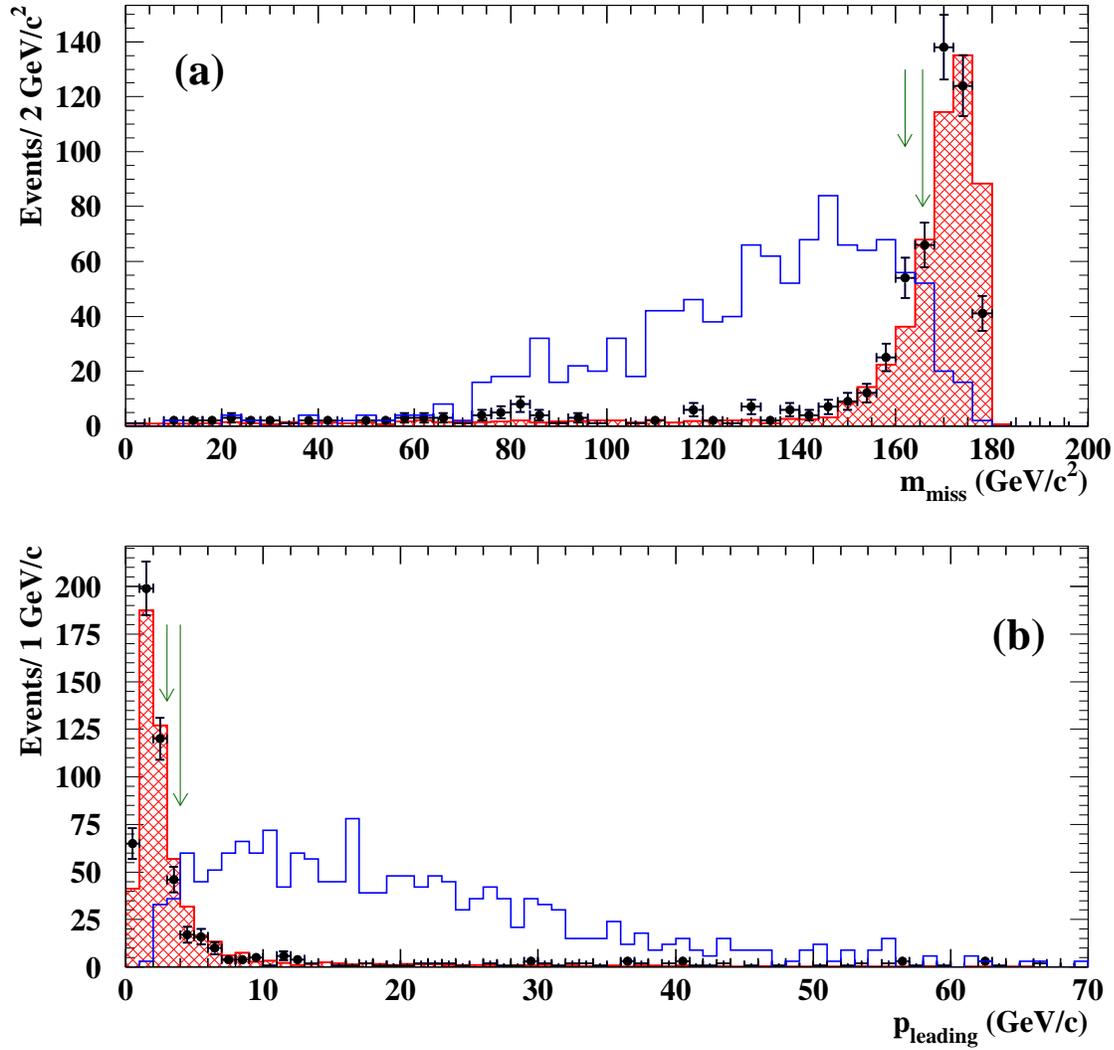}}
\caption[]{(a) missing mass 
and (b) momentum of the leading charged particle,
for data (dots), Standard Model 
simulation (cross-hatched histogram) and one of the simulated signals with 
cross-section not to scale (blank histogram) after preselection 
at $\sqrt{s} = 183$~GeV. 
The arrows indicate selection criteria imposed as explained in the text.}
\label{fig:cut1b}
\end{figure}

\begin{figure}[htbp]\centering
\epsfxsize=16.0cm
\centerline{\epsffile{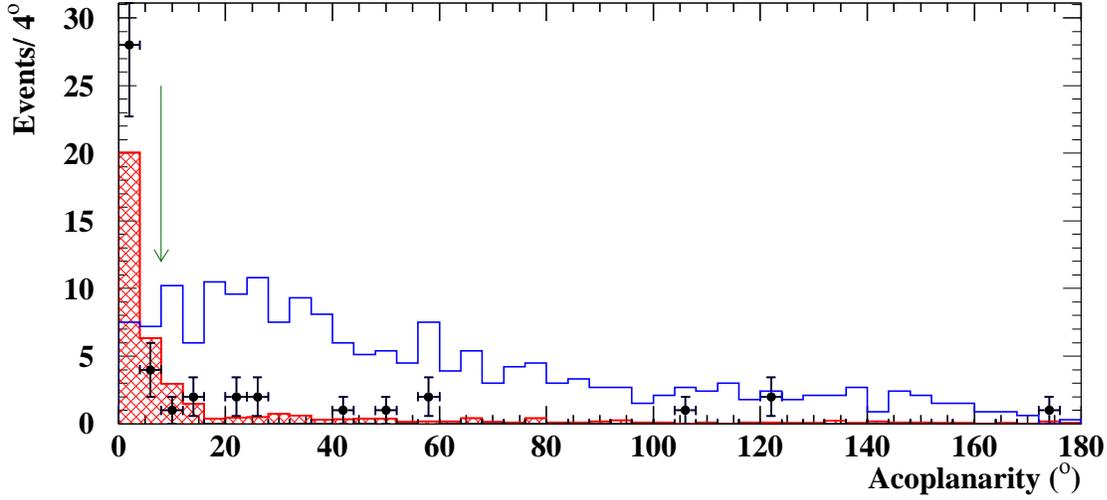}}
\vspace{-8cm}
\caption[]{ Acoplanarity of data (dots),
Standard Model background simulation 
(cross-hatched histogram) and one of the simulated signals with 
cross-section not to scale (blank histogram) at $\sqrt{s} = 183$~GeV,
after the cut to remove 
$\gamma\gamma$\ events. 
The arrow indicates selection criterion imposed as explained in the text.}
\label{fig:cut2a}
\end{figure}

\begin{figure}[htbp]\centering
\epsfxsize=16.0cm
\centerline{\epsffile{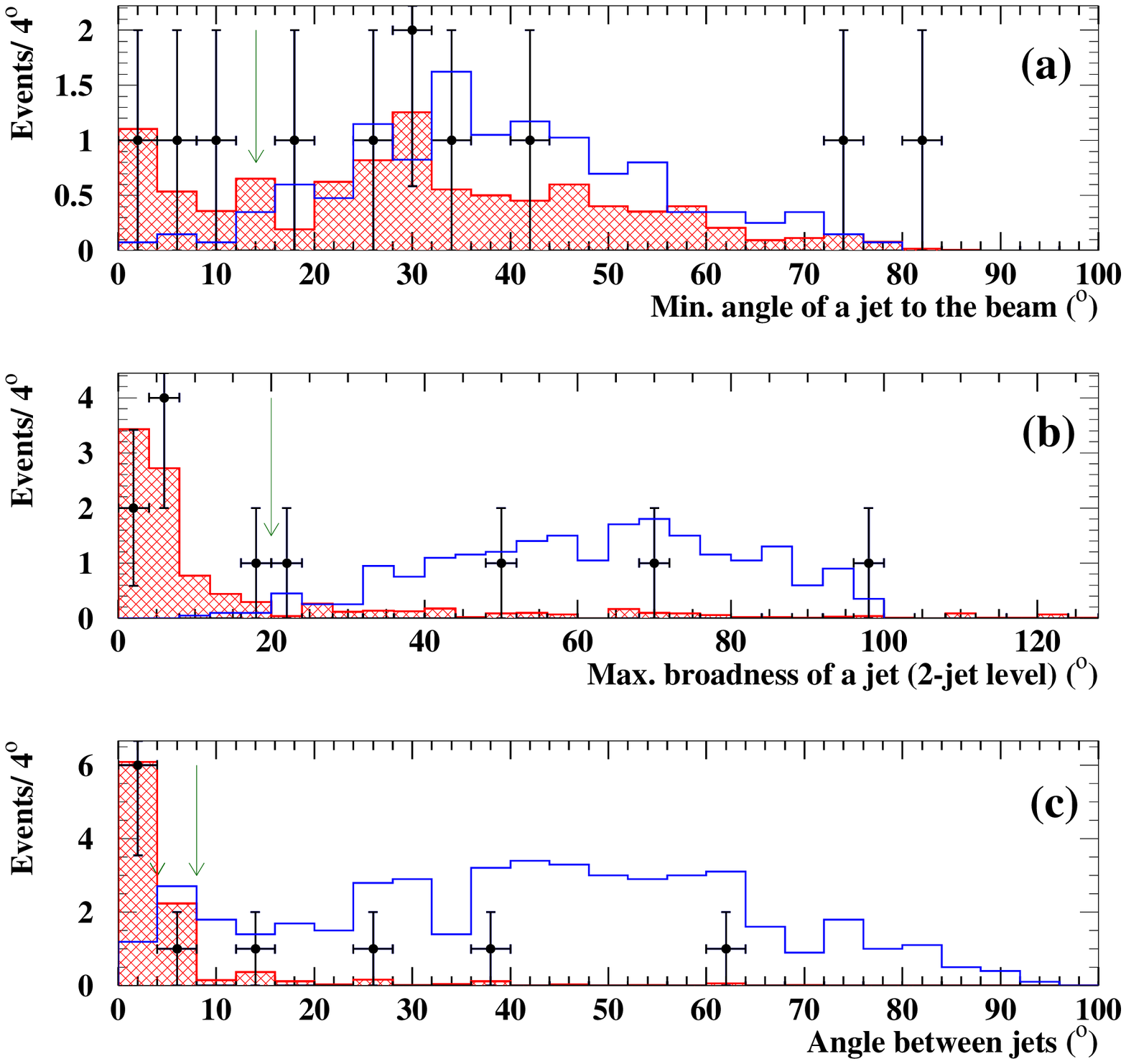}}
\caption[]{(a) minimum angle of a jet to the beam, 
(b) maximum angular broadness of a jet at the 2-jet level and
(c) angle between jets, for data (dots), Standard Model background simulation 
(cross-hatched histogram) and one of the simulated signals with 
cross-section not to scale (blank histogram) at $\sqrt{s} = 183$~GeV,
 after the cut to remove 
$f\bar{f}(\gamma)$\ events. 
The arrows indicate selection criteria imposed as explained in the text.}
\label{fig:cut3}
\end{figure}

%As an example,
Figures~\ref{fig:cut1b} to~\ref{fig:cut3} show some of the
distributions relevant for these selection criteria at $\sqrt{s} = 183$~\GeV.
Table~\ref{tab:evol} shows the effect of these cuts 
at $\sqrt{s} = 183$~\GeV\ on the data, expected 
background and the  
signal for $m_{\tilde \chi^0_1}$ = 75~\GeVcc\ and 
$m_{\tilde \tau_1}$ =  55~\GeVcc. The discrepancy between data and simulation 
on the last bin of figure~\ref{fig:cut1b}-a is attributed to the poor 
description of $\gamma\gamma$ events in the simulation.
%due to the simulation of 
%$\gamma\gamma$-physics events.

\begin{table}[hbtp]
  \begin{center}
    \begin{tabular}{||c||c|c|c|c||c||c||}
      \hline
 Cut & $\gamma\gamma$ &$f\bar{f}\gamma$ & 4-fermion& Total MC & Data
 &                       Signal \\
      \hline
pre-selection  &$496\pm 16 $&$44.5\pm 1.5$&$13.1\pm0.6$&$554\pm16$&567&61.4\%\\
1     &$18\pm 2   $&$40.6\pm 1.4$&$12.1\pm0.6$ &$70.8\pm2.6$&84 &59.2\%\\
2     &$2.2\pm 0.6$&$2.9\pm0.4  $&$4.5\pm0.4 $ &$9.6\pm0.8$ &12 &45.4\%\\
3     & 0          &$0.23\pm0.09$&$0.27\pm0.07$&$0.50\pm0.11$&2 &38.3\%\\
      \hline
    \end{tabular}
  \end{center}
  \caption[]{Number of events remaining in the data and simulated samples
at $\sqrt{s} = 183$~\GeV\ after various stages of the selection procedure
described in the text. The signal efficiencies corresponds to  
$m_{\tilde{\chi}^0_1}=75~$\GeVcc\ and 
$m_{\tilde{\tau}_1}=55~$\GeVcc.}
  \label{tab:evol}
  \end{table}
%%%%%%%%%%%%%%%%%%%%%%%%%%%%%%
%\begin{figure}[htbp]\centering
%\epsfxsize=16.0cm
%\centerline{\epsffile{cut1_1.eps}}
%\caption{Transverse Energy (a) and 
%Energy outside a cone of $30^{\rm o}$ over the total visible energy (b) 
%for data (dots), SM 
%MC (cross-hatched histogram) and one of the simulated signals with 
%cross section not to scale (blank histogram) after 
%preselection. Cuts 
%for these variables
%are explained in the text and are shown with arrows. }
%\label{fig:cut1a}
%\end{figure}
%%%%%%%%%%%%%%%%%%%%%%%%%%%%%%
%%%%%%%%%%%%%%%%%%%%%%%%%%%%%%
%%%%%%%%%%%%%%%%%%%%%%%%%%%%%%
%\begin{figure}[htbp]\centering
%\epsfxsize=16.0cm
%\centerline{\epsffile{cut2_2.eps}}
%\caption{Acoplanarity (c) and Missing Mass (d) for data 
%(dots), SM 
%MC (cross-hatched histogram) and one of the simulated signals with 
%cross section not to scale (blank histogram), 
%after the cut to remove 
%$\gamma\gamma$\ events. Cuts for these variables
%are explained in the text and are shown with arrows. }
%\label{fig:cut2b}
%\end{figure}
%%%%%%%%%%%%%%%%%%%%%%%%%%%%%%
%%%%%%%%%%%%%%%%%%%%%%%%%%%%%%

After these cuts, an efficiency between 25 and 45\% was obtained for 
the signal events, and estimated backgrounds of $0.12  \pm 0.08$, 
$0.15  \pm 0.09$
and $0.50 \pm 0.11$ events at $\sqrt{s}$ = 161, 172 and 183 GeV, respectively.

\subsection{Stau pair production}
This section describes the selection criteria used in the search 
for the process
$e^+ e^- \to \tilde{\tau}_1 \tilde{\tau}_1 \to \tau \tilde{G} \tau \tilde{G}$.
As described in section~\ref{sec:intro}, the mean life-time of the 
$\tilde{\tau}_1$\ depends on the mass of the gravitino. Thus, for a gravitino 
with a mass of the order of a few hundred eV/c$^2$\ or more, the stau would
be sufficiently long lived to decay outside the detector.
When the mass of the gravitino is 
between a few eV/c$^2$\ and a few hundred eV/c$^2$, one or both staus would 
decay in flight in some part of the detector, creating a well defined 
secondary vertex. The search for these decays is described 
in subsection~\ref{searchkinks}.
If the mass of the gravitino is even smaller, stau pair production would
produce displaced vertices. This search is described in 
subsections~\ref{searchimpact} and~\ref{smallimpact}.

\subsubsection{Search for secondary vertices}
\label{searchkinks}
\hspace{\parindent}

This analysis exploits a peculiarity of the
$\tilde{\tau}_1 \rightarrow \tau \tilde{G}$ topology in the 
case of intermediate mass gravitinos, namely,
one or two tracks coming from the interaction point and at least one
of them with either a secondary
vertex or a kink. 
%Reconstruction of secondary vertices is illustrated
%in figure~\ref{fig:grav:def}, which shows a decay vertex and
%the variables used in the analysis.
%\begin{figure}[hbpt]
%\centerline{\epsfxsize=7.0cm \epsfysize=7.0cm \epsfbox{fig-grav-def.eps}}
%  \caption[]{
%    Sketch 
%    illustrating the reconstruction of a secondary vertex,
%    shown in the plane perpendicular to the beam. The stau track 
%   (labelled with $\tilde{\tau}$) and the track of the decay product of the 
%    tau (labelled with $\tau_d$) are extrapolated
%   (dashed line). The extrapolated tracks define a crossing point at 
%   radius $R_{cross}$.
%   $R_{sp}^{\tilde{\tau}}$ and $R_{end}^{\tilde{\tau}}$ 
%   are the radii of the first and last measured points of the 
%   $\tilde{\tau}$ track.
%   $R_{sp}^{\tau_d}$ is the first measured point of track selected as 
%   the $\tau$ decay product. All the radii are measured with respect to 
%   the beam spot (BS).
%} 
%  \label{fig:grav:def}
%\end{figure}

Rather loose preselection cuts, similar to those presented in 
reference~\cite{ref:flying}
were imposed on the events in order to suppress the low energy background
(beam-gas, beam-wall, etc), $\gamma \gamma$, $e^+e^-$ and hadronic events.
To compute the following quantities the reconstructed tracks of charged 
particles were 
required to have momenta above 100~\MeVc\ and impact parameters below 4~cm
in the transverse plane and below 10~cm in the lon\-gi\-tu\-di\-nal
direction. 
Clusters in the calorimeters were interpreted as neutral
particles if they were not associated to charged particles and if their
energy exceeded 100~\MeV. 
%To compute the above quantities only preselected charged and neutral
%particles, described in reference~\cite{chargi}, were used. 
However, no quality requirements were
imposed on the reconstructed tracks in the fo\-llo\-wing  stages.
 
\begin{itemize}
  \item Charged particle multiplicity between 1 and 10;
  \item visible energy above 10~\GeV;
  \item total electromagnetic energy below 40~\GeV;
  \item transverse momentum, computed as the transverse component with respect 
  to the beam axis of the 
  vector sum of the momenta of charged and neutral particles, ${\rm p_T}$, 
  greater than 5~\GeVc;
  \item energy measured in the very forward calorimeters below 10~\GeV.
\end{itemize}
These preselection cuts leave about 0.6\% of the whole data sample.

The  tracks of the events that survived the preselection cuts were
grouped in clusters according to their first measured point (starting
point). This clustering procedure is described in \cite{nhl}.
Each cluster contained all tracks whose starting points differ by less
than 2~cm. The starting point of a cluster was defined as the
average
of the starting points of its tracks. This procedure allowed for
clusters with
a single track if its momentum was larger than 1.5~\GeVc. 
Events were rejected if more than 6 tracks were not grouped in clusters or
no cluster was found.
A cluster with only one track was considered a $\tilde{\tau}_1$
candidate track if its trajectory was compatible with that of a particle
coming from the interaction point (according to the selection criteria
described in reference~\cite{ref:flying}) and its
momentum  was greater than 2~\GeVc.
%:
%\begin{itemize}
%  \item the distance of the first measured point to the beam spot in the
%    plane transverse to the beam axis 
%    ({\it xy} plane), $R_{sp}^{\tilde{\tau}}$, was
%    smaller than 10 cm,
%  \item the momentum of the particle was greater than 2~\GeVc,
%  \item the polar angle of the track respect to the beam axis ($\theta$)
%    $|\cos\theta|<0.8$,
%  \item the impact parameter of the track along the beam axis and in
%    the plane per\-pen\-di\-cu\-lar 
%    to it were less than 10 and 4~cm, respectively.
%\end{itemize}

For each such $\tilde{\tau}_1$ candidate, 
%(single track cluster fulfilling the above conditions)
a search was
made for a second cluster with starting point radius in the transverse plane
({\it xy} plane) greater than 
that of the first measured point 
of the track of the $\tilde{\tau}_1$ candidate, 
%$R^{sp}_{\tilde{\tau}}$ 
and an angular 
separation between the directions defined by the beam spot and the cluster 
starting points smaller than 90$^\circ$ in the {\it xy} plane. This second
cluster was assumed to be
formed by the decay products of the $\tau$ coming from the $\tilde{\tau}_1
\rightarrow \tau \tilde{G}$ process.  Therefore, the $\tilde{\tau}_1$ candidate
and the $\tau$ cluster had to define a secondary vertex. If the $\tau$ cluster
included more than one charged particle, only the one with the highest momentum was
used to
search for the decay vertex or kink (crossing point with
the $\tilde{\tau}_1$ track).

The tracks were parametrised with respect to their perigee~\cite{perig}
to calculate the point of closest approach between the two tracks
(the candidate $\tilde{\tau}_1$ track
and the selected track from the candidate $\tau$\ cluster).
The conditions to define a good crossing point between the
track of the $\tilde{\tau}_1$ and the selected track of the 
$\tau$ decay candidates
are described in
reference~\cite{ref:flying}.

\begin{figure}[hbpt]
%\centerline{\epsfxsize=10.0cm \epsfysize=7.0cm \epsfbox{hadr_183_log.eps}} 
%\centerline{\epsfxsize=10.0cm \epsfysize=7.0cm \epsfbox{show_183_log.eps}} 
%\centerline{\epsfxsize=10.0cm \epsfysize=7.0cm \epsfbox{ang_183_log.eps}} 
\centerline{\epsfxsize=16.0cm \epsfysize=16.0cm \epsfbox{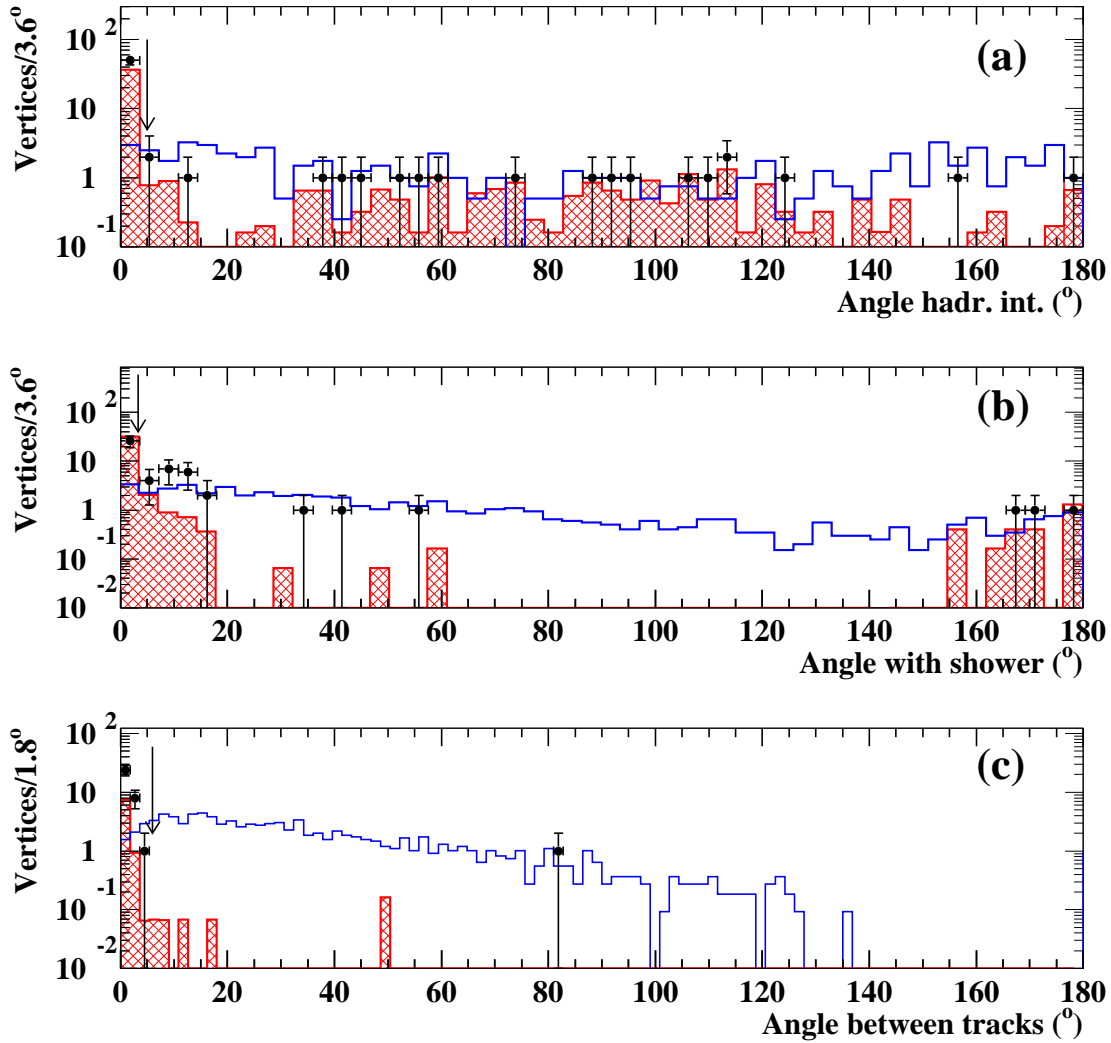}}
 
   \caption[]{(a) angle between the hadronic interaction and the
   reconstructed vertex, (b) angle between the electromagnetic shower
  and the direction defined by the difference between the momenta of 
   $\tilde{\tau}_1$ and its associated $\tau$, defined at the crossing point, 
   and (c) angle between the tracks of the kink,
 for real data (dots), expected Standard Model background (cross-hatched 
histogram) and simulated signal for $m_{\tilde{\tau}_1} = 60$~\GeVcc\
decaying with a mean 
 distance of 50~cm (blank histogram). 
The arrows indicate selection criteria imposed as explained in the text.}
  \label{fig:grav:kinks_BG}
\end{figure}

Fake decay vertices could be present amongst the reconstructed secondary 
vertices, being produced by particles interacting in 
the detector material or by radiated photons if the particle trajectory
was reconstructed into two separated tracks.
To eliminate these classes of events, additional requirements were imposed:
\begin{itemize}
  \item to reject hadronic interactions,
        any reconstructed hadronic interaction (secondary vertices 
        reconstructed in region where there is material) 
        must be outside a cone of half angle 5$^\circ$ around the kink 
        direction; 
  \item to reject photon radiation
        in the case of $\tau$ clusters with only one track,
        there had to be no neutral particle  in a 3$^\circ$ cone
        around the direction defined by the difference between the
        $\tilde{\tau}_1$ momentum and the momentum of the
        $\tau$ daughter calculated at the crossing point; 
  \item to reject segmented tracks,
        the angle between the tracks used to define a vertex
        had to be larger than 6$^\circ$.
\end{itemize}
If no pair of tracks was found to survive these conditions, the
event was rejected.
Fi\-gu\-re~\ref{fig:grav:kinks_BG} shows the distribution of these three angles
for real data, expected Standard Model background 
simulation and simulated signal 
for $m_{\tilde{\tau}_1} = 60$~\GeVcc\ decaying with a mean distance of 50~cm.   
The excess of data in the first bins of figure~\ref{fig:grav:kinks_BG} (c)
is due to underestimation in the simulation of mismatchings between the 
tracking devices.

One event in real data was found to satisfy all the conditions described
above. The event was the superposition of a low energy event with a 
cosmic muon crossing the detector.
However, the two tracks of the cosmic muon follow the cosmic muon rejection
criteria used on subsection~\ref{searchimpact} based on impact parameter and 
on acollinearity. 
Thus, the event will not be considered as a candidate.  
This kind of events was not simulated and therefore its removal 
does not affect the 
calculated efficiencies.

%\begin{figure}[hbpt] 
%\centerline{\epsfxsize=10cm \epsfysize=8cm \epsfbox{eff_dv_183.eps}}
%  \caption[]{ 
%           Efficiency of the vertex search versus the decay
%           distance in the {\it xy} plane
%           for a simulated $\tilde{\tau}_1$ of 60~\GeVcc\ at a 
%           centre-of-mass energy of
%           183~\GeV. }
%  \label{fig:grav:eff_kink}
%\end{figure}

The vertex reconstruction was sensitive to decay lengths in the {\it xy} 
plane, R, 
%(see figure~\ref{fig:grav:eff_kink})
between 15~cm and 90~cm. Within this region a vertex was 
reconstructed with an efficiency of $\sim$54\% since the VD 
(Vertex Detector) and the ID (Inner Detector)  
were needed to reconstruct the
$\tilde{\tau}_1$ track and the TPC (Time Projection Chamber)
to reconstruct the decay products.
The efficiency is flat inside the sensitive region and drops
to zero for
$\tilde{\tau}_1$'s decaying near the outer surface of the TPC. 
%The maximum
%efficiency ($\sim$79\%) was reached when the two staus decayed within this
%sensitive region.
The shape of the efficiency
distribution was independent of the $\tilde{\tau}_1$ mass; it
simply scaled down near the kinematic limit.
The loss of efficiency near the
kinematic limit is due to the fact that the
 $\tilde{\tau}_1$ boost is smaller and the vertex reconstruction less efficient
when the angles between the $\tilde{\tau}_1$ and the $\tau$ products increase. 

The efficiencies for different mean decay lengths and $\tilde{\tau}_1$ masses
were calculated by applying the above selections to the simulated signal
samples.
%Figure~\ref{fig:grav:eff} shows the efficiencies ($\varepsilon_2$) obtained
%for a 60~\GeVcc~$\tilde{\tau}_1$ at a centre-of-mass energy of 183~\GeV\ and
%for mean decay lengths from 0.25 to 1000~cm. For a mean decay length of
%50~cm the vertex search efficiency is of the order of 55\%. 
For a 60~\GeVcc~$\tilde{\tau}_1$ with mean decay length of 50~cm
the vertex search efficiency is of the order of 55\%.

%\begin{figure}[hbpt] 
%\centerline{\epsfxsize=10cm \epsfysize=8cm \epsfbox{eff_l_183.eps}}
%  \caption[]{ 
%           Efficiency of the large impact parameter search ($\varepsilon_1$),
%           vertex search ($\varepsilon_2$), heavy lepton search
%           ($\varepsilon_3$)~\cite{Heavyparticles} and combined  
%            efficiency ($\varepsilon_{tot}$) of these three searches.
%            $\varepsilon_4$\ corresponds to the small impact parameter
%            search. These efficiencies have been calculated from the 
%            simulation of
%            a $\tilde{\tau}_1$ of 60~\GeVcc\ at a centre-of-mass energy of
%            183~\GeV. }
%  \label{fig:grav:eff}
%\end{figure}

\subsubsection{Large impact parameter search}
\label{searchimpact}
\hspace{\parindent}

\begin{figure}[hbpt] 
\centerline{\epsfxsize=12cm \epsfysize=10cm \epsfbox{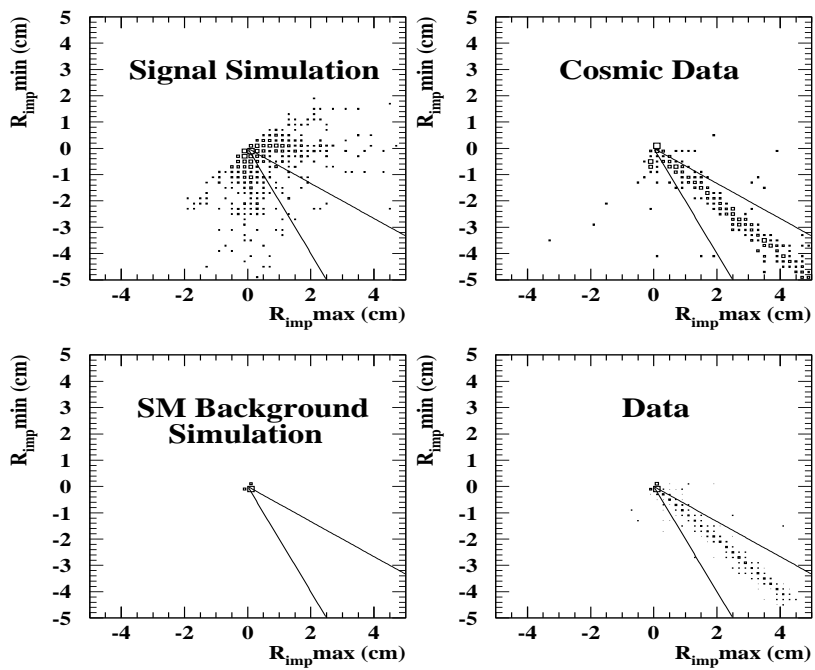}}
  \caption[]{ 
           Impact parameters of two track events for a simulated signal of 
           $m_{\tilde{\tau}_1} = 60$~\GeVcc\ with mean decay length
           of 2.5~cm, cosmic muons, Standard Model expected background 
           and real data. It is plotted the maximum impact parameter versus 
           the minimum one in the $R\phi$ plane. The area between the lines 
           was excluded by the cosmic rejection criteria as described in the 
           text.}
  \label{fig:imp}
\end{figure}

\begin{figure}[htbp]\centering
\epsfxsize=16.0cm
\centerline{\epsffile{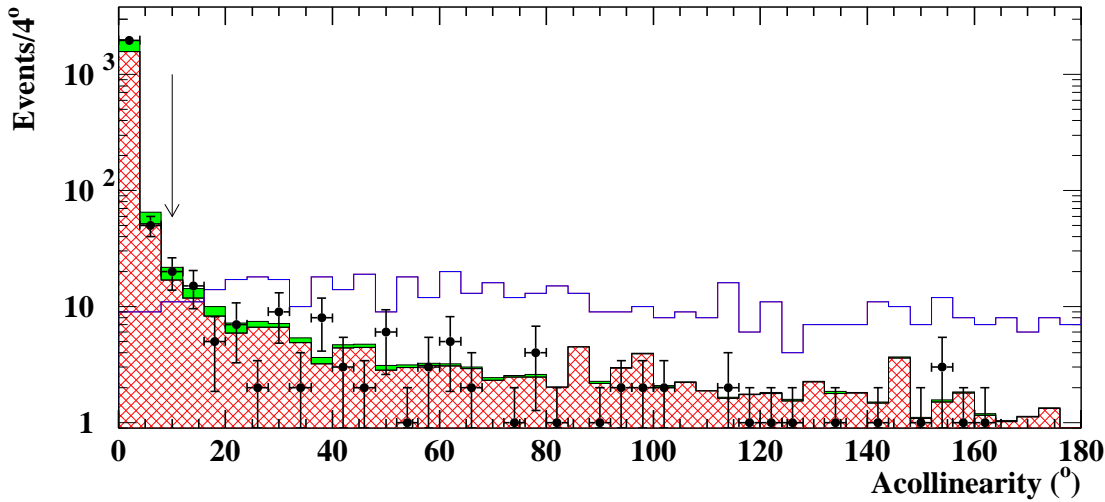}}
\vspace{-8cm}

%\begin{figure}[hbpt]
%\centerline{\epsfxsize=10.0cm \epsfysize=9.0cm \epsfbox{acoll_183.eps}} 
  \caption[]{ Acollinearity 
for real data (dots), a simulated signal 
of $m_{\tilde{\tau}_1} = 60$~\GeVcc\ 
decaying with a mean 
distance of 50~cm (blank histogram), 
and expected simulated Standard Model background 
(cross-hatched histogram) plus cosmic background (dark grey histogram). 
This last background is normalized in order to make the first bin of SM 
background plus cosmic radiation coincide with the corresponding value
of real data.
The selection on this variable is
explained in the text and is shown with an arrow.
} 
  \label{fig:grav:acoll}
\end{figure}

To investigate the region of low gravitino masses (short decay lengths) the
previous search was extended to the case of the $\tilde{\tau}_1$ decaying
between 0.25~cm and around 10~cm. In this case the $\tilde{\tau}_1$
track was not reconstructed in the ID and only the $\tau$
decay products were detected.
The events used
 in this search contained exactly two single track clusters (i.e. two charged
 particles with momentum larger than 1.5 GeV/c and a distance between starting
 points greater than 2 cm) which were acollinear and had large impact
 parameters\footnote{The impact parameter is defined as the distance of 
closest approach of a charged particle to the reconstructed primary vertex. 
The impact parameters in the $R\phi$ and $Rz$ plane are evaluated separately. 
The sign of the impact parameter is defined with respect to the jet
direction. It is positive if the vector joining the primary vertex to the 
point of closest approach of the track is less than 90$^\circ$ from the 
direction of the jet to which the track belongs. In events with two particles,
each reconstructed track is considered as a jet.}.
Cosmic rays, badly reconstructed tracks or interactions in the
detector material could result in large impact parameters. However,
the two tracks in a cosmic event usually have
impact parameters of the same order and opposite sign. The acollinearity
in back-to-back
events with badly reconstructed tracks or interactions was always small.
Figure~\ref{fig:imp} shows the scatter 
plot of the maximum impact parameter versus the minimum one in the 
$R\phi$ plane. Figure~\ref{fig:grav:acoll} shows
the acollinearity distribution for events with two tracks in the TPC. 
Simulated signal events with
$m_{\tilde{\tau}_1} = 60$~\GeVcc\  and a mean decay length
of 2.5~cm are compared with cosmic muon events, simulated Standard Model 
background and real data.
The data points in figure~\ref{fig:grav:acoll} contain cosmic radiation events
 that are not simulated.

%############# NEW

The impact parameter search was only applied to those events accepted by the
same general requirements as in the search for secondary vertices, and not 
selected by the vertex analysis. The events were accepted as candidates if:
\begin{itemize}
% \item there were two single track clusters in the event (i.e. two
%       tracks with momentum larger than 1.5~\GeVc~and a distance between
%       starting points larger than 2~cm);
 \item the first measured point of at least one of the tracks was
       within 12~cm of the beam spot in the plane transverse 
       to the beam axis;
 \item both tracks were reconstructed in the TPC to guarantee a good
       track reconstruction quality;
 \item at least one of the tracks had an impact parameter
       larger than 0.2~cm in the plane transverse
       to the beam axis, to remove
       standard model events;
 \item the ratio of the maximum impact parameter over the minimum impact 
parameter in the $R\phi$ plane  was
          smaller than -1.5 or larger than -0.5, to reject
          cosmic rays;
 \item the acollinearity between the two tracks was larger than 10$^\circ$;
 \item the angle defined by the directions 
       of the starting points of the tracks with respect to the 
       the beam-spot was larger than 3$^\circ$.
\end{itemize}

The efficiencies were derived for the different $\tilde{\tau}_1$ masses and
decay lengths by applying the same selection to
the simulated signal events.
%Figure~\ref{fig:grav:eff} shows
%the efficiency of the large impact parameter search ($\varepsilon_1$) for a
%60~\GeVcc\ $\tilde{\tau}_1$ at a centre-of-mass energy
%of 183~\GeV. For 60~\GeVcc\ $\tilde{\tau}_1$ 
The maximum efficiency was 29.2\%, corresponding to a mean decay
length of 2.5~cm, decreasing very fast for lower decay lengths due
to the cut on minimum impact parameter. For longer decay lengths, 
the appearance of
reconstructed $\tilde{\tau}_1$ tracks in combination with the 
cut on the maximum amount of charged tracks caused the
efficiency to decrease smoothly. This decrease is compensated by a rising
efficiency in the search for vertices.
No dependence on the  $\tilde{\tau}_1$ mass was 
found far from the kinematic limit.
The losses of efficiency for $\tilde{\tau}$ masses near
the kinematic limit and due to initial state radiation were also
considered.\\[0.2cm]

Trigger efficiencies were studied, simulating the DELPHI trigger
response to the events selected by the vertex search and by the large impact
parameter analysis, and were found to be around 99\%.

No events in the real data sample were selected with the above criteria.
The number of expected background events at $\sqrt{s}=183$~GeV is shown in
Table~\ref{tab:grav:bkg} for the combination of the vertex and large
impact parameter searches.

\noindent
\begin{table}[hbt]
\begin{center}
\begin{tabular}{||c|c||} \hline 
%Channel:     & $\tilde{\tau}_1\rightarrow \tau \tilde{G}$     \\ \hline
Observed events
             & 0                                            \\ \hline
Total background
     & 0.63$^{+0.55}_{-0.12}$                              \\ \hline    
\hline
$Z^*/\gamma \to (\tau \tau) (n\gamma)$
     & 0.07$^{+0.16}_{-0.06}$ \\ \hline
$Z^*/\gamma \to (ee) (n\gamma)$
   & 0.09$^{+0.19}_{-0.03}$ \\ \hline
4-fermion (except $\gamma\gamma$)
   & 0.10$^{+0.12}_{-0.06}$\\ \hline
$\gamma\gamma \to \tau^+\tau^-$
    & 0.20$^{+0.27}_{-0.06}$\\ \hline
$\gamma\gamma \to e^+e^-$
    & 0.17$^{+0.39}_{-0.05}$ \\ \hline
\end{tabular}
\end{center}
\caption[.]{
The number of observed events at $\sqrt{s}=183$~GeV,
together with the total number of expected background events
and the expected numbers from the individual background sources,
for both large impact parameter and secondary vertex searches combined.}
\label{tab:grav:bkg}
\end{table}

%\newpage
\subsubsection{Small impact parameter search}
\label{smallimpact}
\hspace{\parindent}

The large impact parameter search can be extended further down to 
mean decay lengths of around 0.1~cm.
%The criteria used to select events with short lifetime
%were defined on the basis of
%the simulated signal and background events. 
Charged particles were selected
if their impact parameter was less than 10 cm in the transverse plane and
less than 15 cm in the longitudinal direction and their polar angle between 
20$^{\circ}$~and 160$^{\circ}$. 
Their measured momentum was required to be larger than 400~\MeVc\,
with relative error less than 100\% and track length
larger than 30 cm. Any calorimetric deposit 
associated to a discarded charged particle was 
assumed to come from a neutral particle.

This search was restricted to events with 2 to 4 charged particles
and missing energy larger than 0.3 $\sqrt{s}$.  
The $\gamma\gamma$ events were suppressed by requiring that the 
visible energy (E$_{vis}$) be bigger than
0.08 $\sqrt{s}$ 
and the transverse missing momentum larger than 0.03 $\sqrt{s}$.
The polar angle of the missing momentum was
required to be between  30$^{\circ}$\ and 150$^{\circ}$\
and the total energy in the forward and backward regions (${\mathrm E}_{30}$) 
was required to be less than 10\% of the total visible 
energy in the event.

The events were then divided into two hemispheres using the thrust axis.
The total momentum of charged and neutral particles 
in each hemisphere was computed and used to define the events' acollinearity.
Standard $e^+e^- \rightarrow f \bar f (\gamma)$ 
processes and cosmic ray events were reduced by
requiring the acollinearity to be greater than 10$^{\circ}$.
The charged particle with largest momentum
in each hemisphere was selected (leading particle).
%These two selected charged particles are called in the following 
%leading particles and referred to with subscripts 1 and 2.
The following quality requirements were only applied to the leading 
charged particles: the first measured point of the tracks had to be within 
50 cm of the beam spot in the {\it xy} plane, 
the tracks were required to have at least a track segment beyond the ID
detector, and  
away from insensitive regions of the electromagnetic calorimeter.
In addition, at least one of the tracks was required to be 
reconstructed with the TPC.

The standard $e^+e^- \rightarrow f \bar f (\gamma)$ and cosmic
backgrounds were reduced by requiring the angle between the leading
particles in the {\it xy} plane to be 
less than 3 radians. 
$q \bar{q}(\gamma)$ and 4-fermion 
events were further rejected by requiring $\sqrt{p_1^2+p_2^2}$ (where $p_1$ 
and $p_2$ are
the momenta of the leading particles) to be smaller than  $0.03 \sqrt{s}$.
To reduce Bhabha events the total electromagnetic energy of the leading
particles (E$_{em1}$ + E$_{em2}$)
had to be smaller than $0.35 \sqrt{s}$. 
By requiring 
that any leading track with an impact parameter larger than 1 cm in 
the  {\it xy} plane should be reconstructed by the TPC and at least another 
detector, the residual cosmics were rejected.
Finally, $\gamma$ conversion events with only 
two tracks
were rejected by requiring
the angle between tracks at their perigee to be greater than 1$^{\circ}$.

\begin{figure}[htbp]
\centerline{\hbox{
\psfig{figure=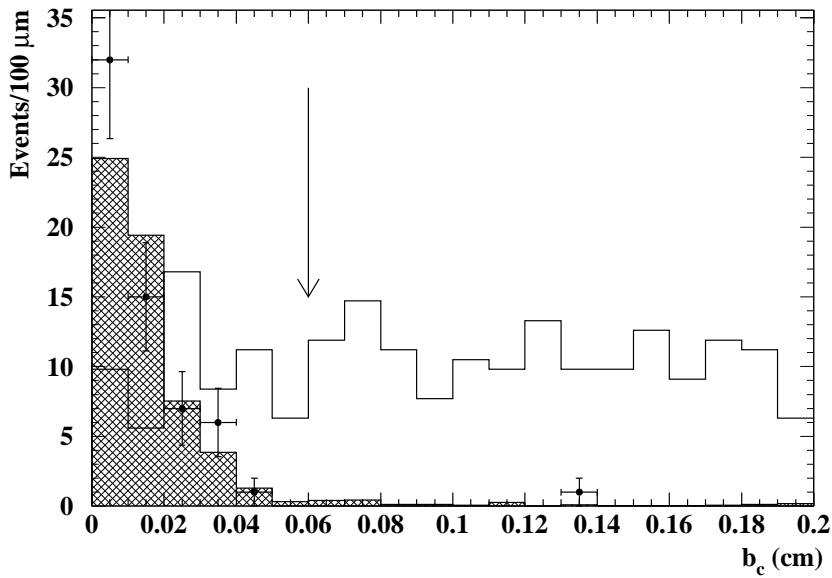,width=12 cm}}}
\caption[]{$b_c$ distribution 
for real data (dots), expected Standard Model background 
(cross-hatched histogram) 
and a simulated signal of $m_{\tilde{\tau}_1} = 50$~\GeVcc\  
and a mean decay length of $\sim$ 1 cm (in arbitrary scale). }
  \label{cavallo1}
\end{figure}

The background left after the selection described above consists
mainly of events containing $\tau$ pairs in the final state
($\gamma^*/ Z^* \rightarrow \tau \tau$ and 
$WW \rightarrow \tau \nu\tau\nu$).
To reject these events, the variable 
$b_c = \sqrt{b_1^2+b_2^2}$ was used, where $b_1$ and $b_2$ are 
the impact parameters of the leading particles,
defined in the  {\it xy} plane. 
Figure \ref{cavallo1} shows the $b_c$
distribution of the selected real data, the total residual simulated 
background and the $b_c$ distribution of one set of simulated signal events, 
with an arbitrary scale.
A cut of $b_c \ge 600 \mu$m was chosen to reject most of the remaining
background.

Applying these cuts to the simulated signal events, 
the efficiency turned out not to 
depend separately on the centre-of-mass energy and
on the $\tilde{\tau}_1$ mass but rather on the $\tilde{\tau}_1$ decay 
length in the
laboratory system, which is determined by both these variables. 
The maximum efficiency was $\sim$ 40\% for a mean decay length of 
$\sim$ 2~cm. The cut on $b_c$ caused the efficiency to drop 
at small decay lengths ($\sim$ 15\% at 1~mm), whilst at large decay 
lengths a loss of efficiency 
was due to the upper impact parameter cut used in the track selection.

%\begin{figure}[htbp]
%\centerline{\hbox{
%\psfig{figure=effi_all.eps,height=10 cm}}}
%\caption{ Selection efficiency as a function of the $\tilde{\tau}_1$ 
%decay length
%in the lab. system }
%  \label{cavallo3}
%\end{figure}

 The non-negligible background contributions,
normalized to the integrated luminosities of the four samples, are shown 
in table \ref{tab:cavallo1} together with the number of selected data events. 
As expected, the main sources of background come from channels containing
$\tau$'s in the final state. 

\begin{table}[hbtp]
  \begin{center}
    \begin{tabular}{||c|c|c|c|c||}
      \hline
 &130 GeV + &161 GeV &172 GeV &183 GeV \\
 & 136 GeV & & & \\
\hline
 Observed events & $ 0 $ & $ 0 $ &
 $ 0 $  &$ 1 $\\
\hline
 Total background & $ 0.19^{+0.66}_{-0.08}$&$0.33^{+0.19}_{-0.11}$&$0.19^{+0.12}_{-0.05}$
&$1.97^{+0.46}_{-0.27}$\\
\hline \hline
 $Z^*/\gamma \to (\tau \tau) (n\gamma)$  &
  $0.19^{+0.14}_{-0.08}$&$0.21^{+0.13}_{-0.09}$&$0.07^{+0.07}_{-0.04}$&$
      0.99^{+0.29}_{-0.22}$ \\
\hline
 $\gamma \gamma \rightarrow \tau^+ \tau^-$ &
 $0.00^{+0.65}_{-0.00}$&$0.09^{+0.14}_{-0.06}$&$0.00^{+0.08}_{-0.00}$&$
              0.20^{+0.33}_{-0.12}$\\
\hline
 WW                                    &
 -&$0.03^{+0.01}_{-0.01}$&$0.12^{+0.04}_{-0.03}$&$0.76^{+0.13}_{-0.11}$ \\
\hline
 ZZ                                    &
 -&-&-&$0.02^{+0.02}_{-0.01}$\\
\hline
% $ ee\tau\tau $                          &
%   $10^{-3}$&$0.16^{+0.25}_{-0.10}$&$0.00^{+0.05}_{-0.00}$ &$10^{-3}$\\
%\hline
    \end{tabular}
  \end{center}
  \caption{Expected simulated SM background events and selected data 
events at the various centre-of-mass energies for the small 
impact parameter search.}
  \label{tab:cavallo1}
  \end{table}

%%%%%%%%%%%%%%%%%%%%%%%%%%%%%%
   
%%%%%%%%%%%%%%%%%%%%%%%%%%%
%%resultados
%%%%%%%%%%%%%%%%%%%%%%%%%%%%
\section{Results and interpretation}
\label{sec:resultados}
\subsection{Neutralino pair production}
\label{sec:resultados:neutralino}
\begin{figure}[htbp]\centering
\epsfxsize=10.0cm
\centerline{\epsffile{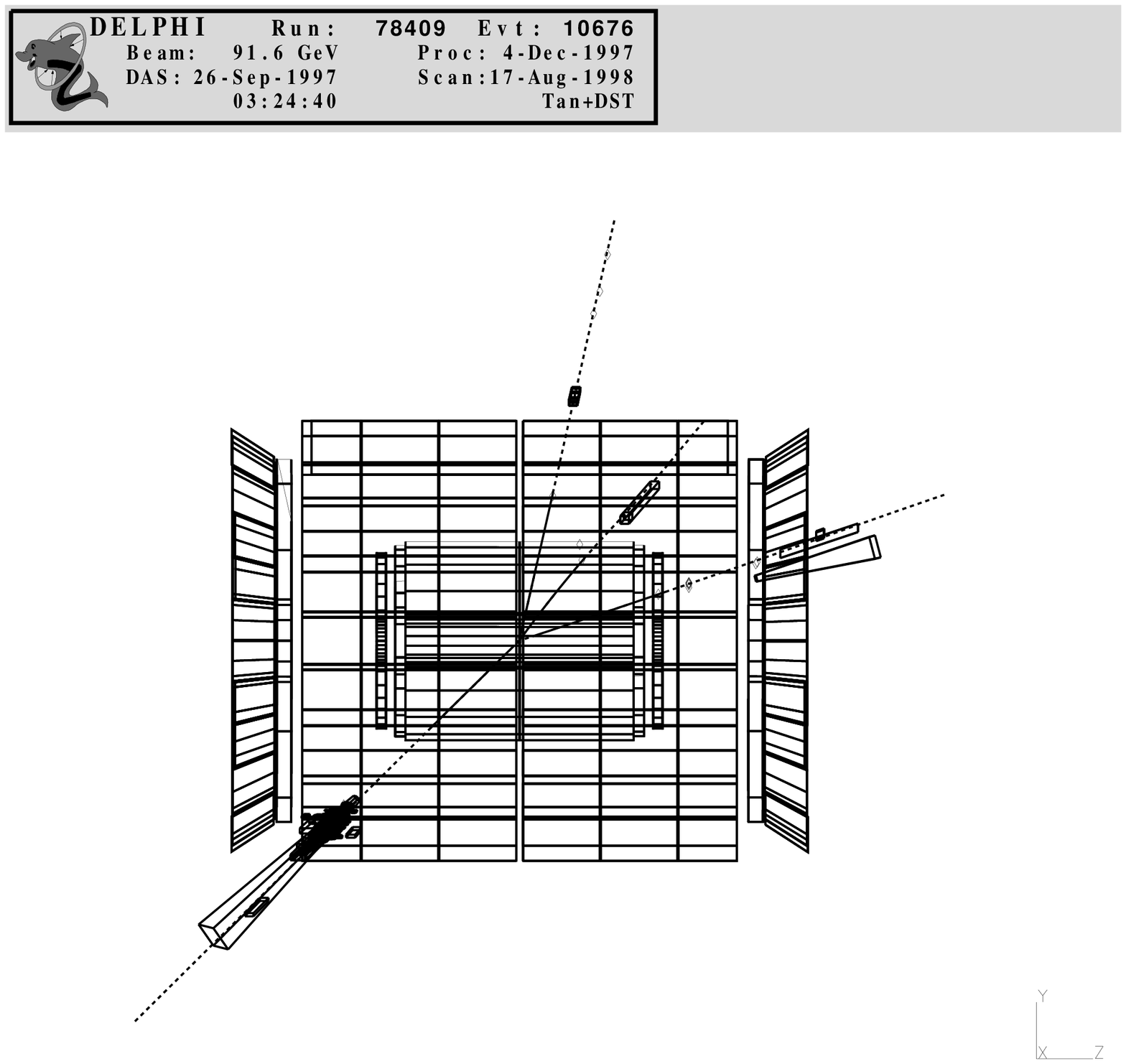}}
\vspace{-1.0cm}
\caption[]{{\it yz} view of the second of the two neutralino
pair production candidates. From above, clockwise, 
the particles are identified as a muon, a pion, an electron 
(with associated radiated photon) and an electron.
}
\label{fig:can2b}
\end{figure}

No event passes the selections at $\sqrt{s}=161$\ or 172~\GeV.
Two events were observed to pass all the cuts at $\sqrt{s}=183$~\GeV.
One of them is shown in fig.~\ref{fig:can2b}.

Their main features are listed in Table~\ref{tab:cands}.
Both of them can be interpreted as being 4-fermion events.
The first event has an electron, a pion, and two unidentified 
low momentum particles. 
The event could be described as
$\gamma^* \gamma^*$, each virtual photon  going into a pair of $\tau$s.
The second event contains a muon, two energetic electrons and a pion.
It could be described as a $Z^* \gamma^*$\ event, with $Z^* 
\rightarrow e^+ e^-$ and $\gamma^* \rightarrow \tau^+\tau^-$.

%%%%%%%%%%%%%%%%%%%%%%%%%%%%%%
\begin{table}[hbtp]
  \begin{center}
    \begin{tabular}{||c|c|c||}
      \hline
  & Candidate 1 & Candidate 2 \\
      \hline
$r$ & 0.14 & 0.31\\
$p_{\rm T}$ &8.7~\GeVc & 9.2~\GeVc \\
$m_{miss} $ &139.5~\GeVcc & 63.8~\GeVcc \\
Thrust & 0.91& 0.84\\
${\rm E}_{30}/{\rm E}_{vis}$ & 0.45& 0.62 \\
${\rm E}_{T}$ & 28.0~\GeV & 78.6~\GeV \\
Acoplanarity & $8.6^{\circ}$ & $15.9^{\circ}$ \\
%Num. of charged & 4& 6\\
%tracks & & \\
Num. of charged particles  & 4& 6\\ 
%Min. angle  & $63^{\circ}$&$26.^{\circ}$ \\
%between jets & & \\
Min. angle between jets & $63^{\circ}$&$26.^{\circ}$ \\
%P of leading & 17.8& 43.7 \\
%track & & \\
P of leading particle & 17.8~\GeVc & 43.7~\GeVc  \\

      \hline
    \end{tabular}
  \end{center}
  \caption[]{Some characteristics of the two candidates found at 
$\sqrt{s} = 183$~\GeV.}
  \label{tab:cands}
  \end{table}
%%%%%%%%%%%%%%%%%%%%%%%%%%%%%%
%\begin{figure}[htbp]\centering
%\epsfxsize=10.0cm
%\centerline{\epsffile{r77798_e3574_xy.psc}}
%\caption{xy view of the first of the two neutralino
%pair production candidates.
%}
%\label{fig:can1a}
%\end{figure}
%%%%%%%%%%%%%%%%%%%%%%%%%%%%%%
%\begin{figure}[htbp]\centering
%\epsfxsize=10.0cm
%\centerline{\epsffile{r077798e003574_zy.ps}}
%\vspace{-1.0cm}
%\caption{yz view of the first of the two neutralino
%pair production candidates decaying into four taus plus missing energy.
%}
%
%\label{fig:can1b}
%\end{figure}
%%%%%%%%%%%%%%%%%%%%%%%%%%%%%%
%\begin{figure}[htbp]\centering
%\epsfxsize=16.0cm
%\centerline{\epsffile{/r78409_e10676_xy.psc}}
%\caption{xy view of the second of the two candidates.
%}
%\label{fig:can2a}
%\end{figure}
%%%%%%%%%%%%%%%%%%%%%%%%%%%%%%
%%%%%%%%%%%%%%%%%%%%%%%%%%%%%%
Since no evidence for a signal was found in the data, 
a limit on the production cross-section for neutralino pairs was derived for
each ($m_{\tilde{\chi}_1^0}$,$m_{\tilde{\tau}_1}$) combination.  
A statistical error of $\pm$1.5\% was assumed for the 
signal efficiency. 

In what follows, the model described in reference~\cite{Dutta} will
be used in order to derive limits. This is a general model which 
assumes only radiatively broken electroweak symmetry and 
null trilinear couplings at the messenger scale. The 
corresponding parameter space was scanned as follows:
$1\leq n \leq 4$, $5\ {\rm TeV}\leq\Lambda\leq 900\ {\rm TeV}$, 
$1.1\leq M/\Lambda \leq 9000$, $1.1\leq \tan\beta\leq 50$, and $\mu>0$,
where
$n$\ is the number of messenger generations in the model, $\Lambda$\ 
is the ratio between the vacuum expectation values 
of the auxiliary component superfield and the scalar component of the
superfield and $M$\ is the messenger mass scale, tan $\beta$\ and $\mu$\ are
defined as for the MSSM.

\begin{figure}[htbp]\centering
\epsfxsize=14.5cm
\centerline{\epsffile{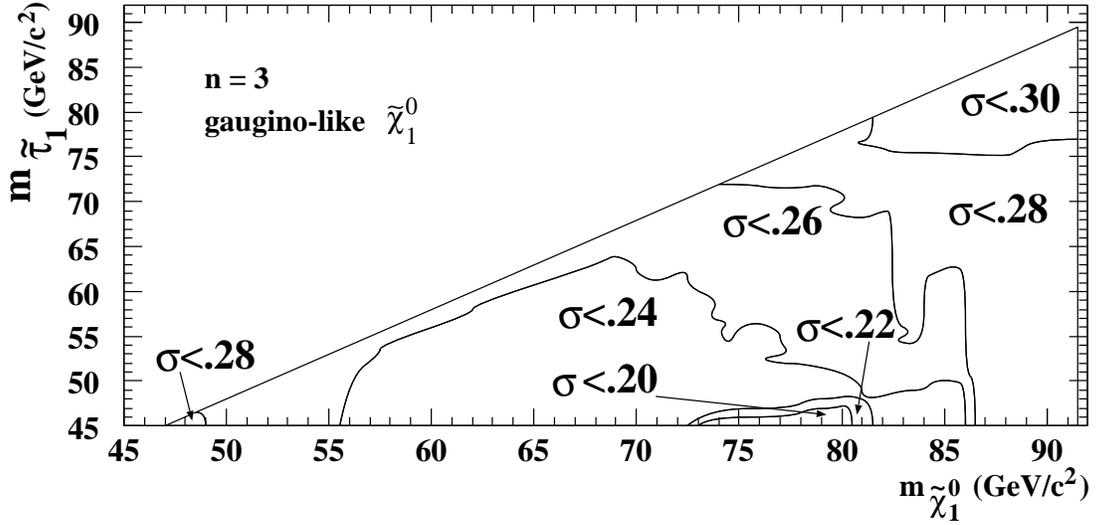}}
\caption[]{95\% C.L. upper limit of the \nuno~pair 
production cross-section (in picobarn) at $\sqrt{s}=18$3~\GeV 
after combining the results of the searches at $\sqrt{s} = 161$, 172 
and 183~\GeV, as a
function of $m_{\tilde{\chi}_1^0}$\ and $m_{\tilde{\tau}_1}$
for the case $n=3$\ and gaugino-like neutralinos, where $n$~ is
the number of messenger generations. The diagonal and vertical lines
show respectively the limits $m_{\tilde{\chi}_1^0} = 
m_{\tilde{\tau}} + m_{\tau}$
and $m_{\tilde{\chi}_1^0} = \sqrt{s}/2$.
%The grey-shaded area is eliminated by~\cite{ref:flying}.
}
\label{fig:limit}
\end{figure}

Figure~\ref{fig:limit} 
shows the 95\% C.L. upper limit on the \nuno~pair 
production cross-section at $\sqrt{s} = 183$~GeV as a
function of $m_{\tilde{\chi}_1^0}$\ and $m_{\tilde{\tau}_1}$\ 
after combining the results of the searches at $\sqrt{s} = 161$, 172 
and 183~\GeV\ with the maximum likelihood ratio method~\cite{Read}.
For different number of messenger generations,
the ratios between production cross sections at different energies
are bound to  vary within certain limits. 
The same happens when 
considering scenarios with higgsino- or gaugino-like neutralinos.
Figure~\ref{fig:limit} presents as an example
the case of $n=3$\ and gaugino-like 
$\tilde{\chi}_1^0$. For the other scenarios considered in this study
($1\leq n\leq 4$, and gaugino- or higgsino-like neutralinos),
the maximum difference with respect to figure~\ref{fig:limit} occurs
in the region where 
$m_{\tilde{\chi}_1^0} <80$~\GeVcc\ and $m_{\tilde{\tau}_1} <65$~\GeVcc, and 
is not bigger than 10\%.

%%%%%%%%%%%%%%%%%%%%%%%%%%%%%%
%%%%%%%%%%%%%%%%%%%%%%%%%%%%%%

Given the aforementioned limits on the production cross-section, 
some sectors of the 
($m_{\tilde{\chi}^0_1},m_{\tilde{\tau}_1}$) space can be excluded.
In order to achieve the maximum sensitivity, the results from 
two other analyses are taken into account. 
The first is the search for $\tilde{\tau}_1$ pair production
in the context of the MSSM. In the case where the MSSM $\tilde{\chi}_1^0$\ is 
massless, the kinematics correspond to the case of $\tilde{\tau}_1$\ 
decaying into a $\tau$\ and a gravitino, except for spin effects, 
which are not taken into account in SUSYGEN.
The second is the search for lightest neutralino pair production 
in the region of the mass space where  
$\tilde{\chi}_1^0$\ is the NLSP~\cite{2gamma} (the region above the diagonal 
line, i.e. $m_{\tilde{\tau}} > m_{\tilde{\chi}^0_1}$). 
Within this zone, the neutralino decays into a gravitino and a photon.

\begin{figure}[htbp]\centering
\epsfxsize=14.5cm
\centerline{\epsffile{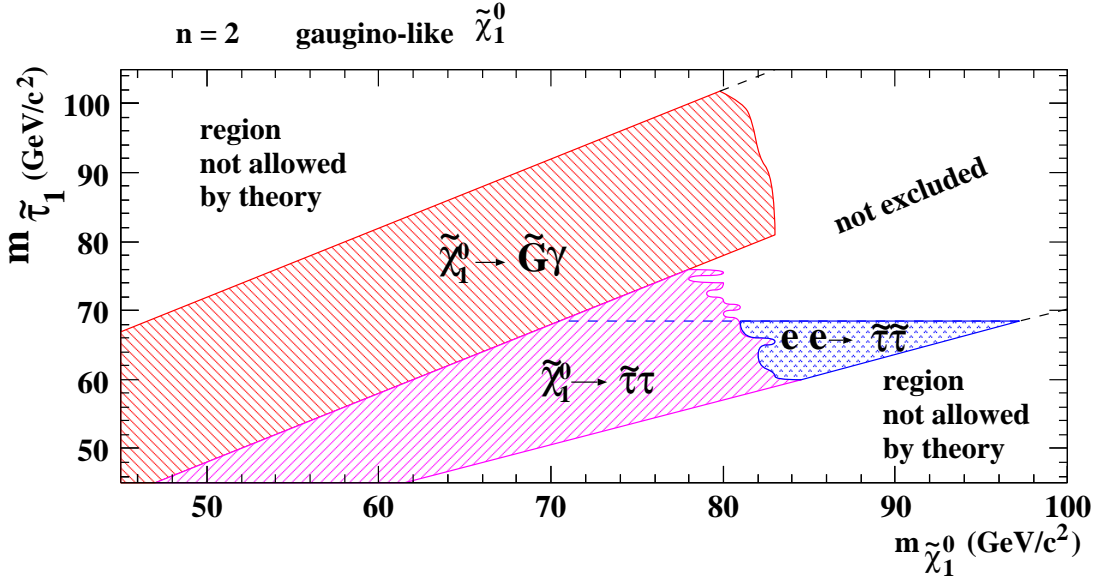}}
\caption[]{Areas excluded at 95\% C.L. for $n=2$, gaugino-like neutralinos
and $m_{\tilde{G}}< 1$~eV in the 
$m_{\tilde{\chi}_1^0}$\ {\it vs.} $m_{\tilde{\tau}_1}$ plane. 
The positive-slope dashed
 area is excluded by this analysis.
The  negative-slope dashed
area is excluded by the search for
$\tilde{\chi}^0_1\rightarrow \gamma \tilde{G}$,
 and the point-hatched area by the direct search for stau pair 
production in the 
MSSM framework. The wiggled curve indicating the limit 
from the search for neutralino pair production, is due to a 
rapid variation in the 95\% C.L. limit on the production
cross section.}
\label{fig:masses}
\end{figure}

As an illustration, fig.~\ref{fig:masses} 
presents the 95\% C.L. excluded areas for the case
$n=2$\  and gaugino-like neutralinos in the 
$m_{\tilde{\chi}_1^0}$\ {\it vs.} $m_{\tilde{\tau}_1}$ plane. 
The positive-slope dashed area 
is excluded by this analysis. The resulting 95\% C.L. 
lower limit
on the mass of the lightest neutralino is 78~\GeVcc.
%The dashed-purple
%line shows the predicted limit~\footnote{The predicted limit is the limit 
%that would obtained if the number of observed data events equals the number of 
%predicted MC background events.}. 
The negative-slope dashed area is excluded by
the analysis searching for neutralino pair production followed by the decay
$\tilde{\chi}^0_1\rightarrow \tilde{G}\gamma$. 
The point-hatched
 area is excluded by the direct search for MSSM $\tilde{\tau}_1$ pair
production~\cite{stau-pair}, taking into account the possibility of
$\tilde{\tau}_L - \tilde{\tau}_R$\ mixing~\cite{bartl}.

For other cases, lower limits 
for the mass of the lightest neutralino obtained with this analysis 
are described in table~\ref{tab:mass}. In the case of $n=1$\ and 
gaugino-like lightest neutralino, the NLSP is always $\tilde{\chi}_1^0$, 
and the lower limit is derived from the search for acoplanar 
photons~\cite{2gamma}.
%%%%%%%%%%%%%%%%%%%%%%%%%%%%%%
\begin{table}[hbtp]
  \begin{center}
    \begin{tabular}{||c|c|c||}
      \hline
 $n$ & gaugino-like $\tilde{\chi}_1^0$ & higgsino-like $\tilde{\chi}_1^0$  \\
     & (\GeVcc ~)                       &            (\GeVcc ~)            \\
      \hline
1 & 81.0 & 71.0\\
2 & 78.0 & 71.0\\
3 & 77.0 & 49.0\\
4 & 78.0 & 45.0\\
      \hline
    \end{tabular}
  \end{center}
  \caption{The 95\% C.L. lower limits on $m_{\tilde{\chi}_1^0}$
for eight different scenarios. When $n=1$\ and the lightest neutralino 
is gaugino-like, the limit comes from the search for two acoplanar photons.}
  \label{tab:mass}
  \end{table}
%%%%%%%%%%%%%%%%%%%%%%%%%%%%%%

%\subsection{Exclusion limits in the (M$_{\tilde{\tau}}$, 
%M$_{\tilde{G}}$) plane}
\subsection{Stau pair production}

%\hspace{\parindent}

%One candidate was observed to pass the cuts of the small impact 
%parameters search. 
No candidate was observed to pass the selection of the large impact 
parameter and secondary vertex searches while the total number of background 
events expected was 0.63 (0.37 on the vertex search and 0.26 on the large
impact parameter search).
The results of these analyses were
combined with those of the stable heavy lepton search described
in~\cite{Heavyparticles}, which considers
$\tilde{\tau}_1$ decays outside the
tracking devices ($R > $ 200~cm).
%Figure~\ref{fig:grav:eff} shows
%the efficiency of the stable heavy lepton search ($\varepsilon_3$)
%for a stau mass of 60~\GeVcc\ at a centre-of-mass energy of 183~\GeV\
%as a function of the decay length.
For very large $\tilde{\tau}_1$ masses, efficiencies
around 80\% were obtained by the heavy lepton search.
%Also shown
%is the combined efficiency of the three analyses
%($\varepsilon_{tot}$), large impact parameter, vertex and heavy 
%lepton analyses. 
Given that an event could be selected both by the vertex search 
and by the stable
heavy lepton search,
the correlation was taken into account. 

One candidate was observed to pass the small impact 
parameters search, due to a leading track with $b_1 = 1.$3 mm caused
by an interaction with the material of the microvertex detector. 
A maximum efficiency of around 40\% was estimated for this search,
and the expected SM background was 1.97 events.

\begin{figure}[htbp]\centering
\epsfxsize=16.0cm
\centerline{\epsffile{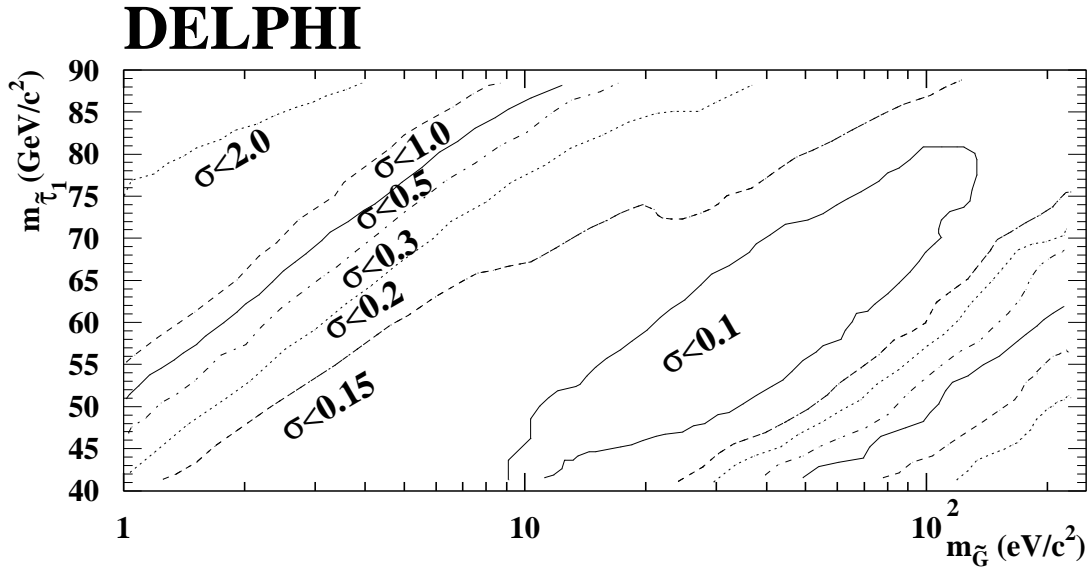}}
\vspace{-7.5cm}

%\begin{figure}[htb]
%\centerline{\epsfxsize=10.0cm \epsfysize=10.0cm \epsfbox
%{cross_comb.eps}}
  \caption[]{
95\% C.L. upper limit of the $e^+e^-\to\tilde{\tau}\tilde{\tau}$\  
production cross-section (in picobarn) at $\sqrt{s}=18$3~\GeV\ 
after combining the results of the searches at $\sqrt{s} = 161$, 172 
and 183~\GeV. Results are shown in the 
($m_{\tilde{G}}$,$m_{\tilde{\tau}_1}$) plane. Searches for events containing
charged tracks  with small impact parameter, large impact 
    parameter or vertices are used.} 
  \label{fig:grav:cross}
\end{figure}

Figure~\ref{fig:grav:cross}
shows the 95\% C.L. upper limit on the stau pair production cross-section 
at $\sqrt{s} = 183$~GeV
after combining the results of the searches at $\sqrt{s} = 161$, 172 
and 183~\GeV\ with the maximum likelihood ratio method~\cite{Read}.
The results are presented in the 
($m_{\tilde{G}}$,$m_{\tilde{\tau}_1}$) plane combining the two impact
parameter searches and the vertex analysis. 
The minimum upper limits achieved for a given $\tilde{\tau}_1$ were around
0.10-0.15~pb depending on $m_{\tilde{G}}$.
For $13~{\mathrm eV}/c^2  < m_{\tilde{G}} < 150~{\mathrm eV}/c^2$ and a 
70.0~\GeVcc\ $\tilde{\tau}_1$, a 0.15~pb limit was obtained.

%The upper cross section limits at 95\%~C.L. on the 
%production of $\tilde{\tau}_1$ for this search 
%are shown in Figure~\ref{fig:grav:cross:fc}.
%For a 70~\GeVcc\ $\tilde{\tau}_1$ and 
%$ 20~{\mathrm eV}/c^2 > m_{\tilde{G}} > 10~{\mathrm eV}/c^2$,
%a 0.2~pb limit was obtained. 

%
%\begin{figure}[htb]
%\centerline{\epsfxsize=10.0cm \epsfysize=10.0cm \epsfbox{sigmaplot.eps}}
%  \caption[]{ 
%    Upper cross section limit at 95\% C.L. in the 
%    ($m_{\tilde{\tau}_1}$, $m_{\tilde{G}}$) plane for 183~\GeV\
%    centre of mass energy obtained from small impact 
%    parameter search.
%} 
%  \label{fig:grav:cross:fc}
%\end{figure}
% 

The upper limits on the production cross-section were
used to exclude $m_{\tilde{\tau}_1}$ values as a function of
$m_{\tilde{G}}$ combining all LEP2 energies, assuming conservatively the
$\tilde{\tau}_1$ to be right-handed. The vertex analysis
allows the exclusion of $\tilde{\tau}_R$ masses between
70.0 and 77.5~\GeVcc\  at 95\%~C.L. in the range of intermediate 
gravitino masses (25 to 150~\eVcc),
the stable heavy lepton search covers the high gravitino mass region
(over 100~\eVcc), while the large and small impact parameter
searches cover the region of low gravitino masses.
  
\begin{figure}[htb]
\centerline{\epsfxsize=10.0cm \epsfysize=10.0cm \epsfbox{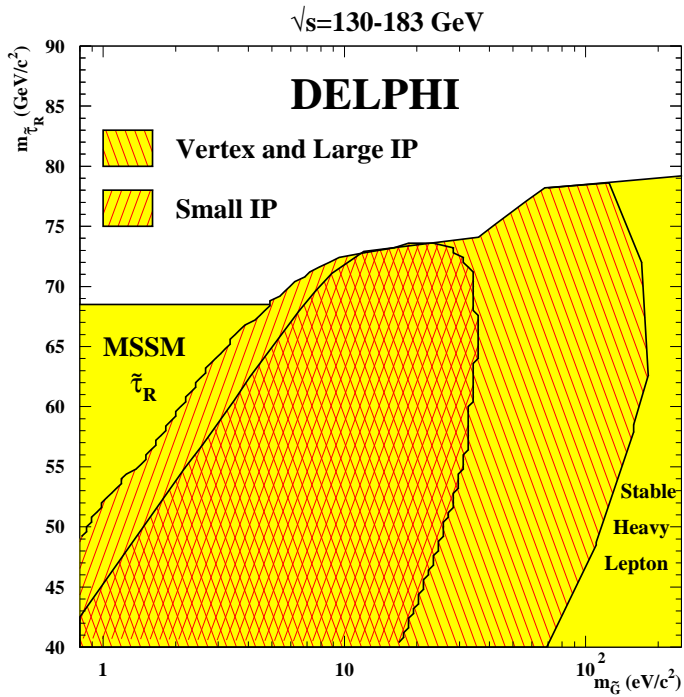}}
  \caption[]{ 
    Exclusion region in the 
    ($m_{\tilde{G}}$,$m_{\tilde{\tau}_R}$) plane
    at 95\%~C.L. for the present analysis combined 
    with the stable heavy
    lepton search and the MSSM $\tilde{\tau}_R$ search, using all LEP-2
    data. 
%    The region 
%    covered by the 
%    LEP I searches of stable heavy leptons is also shown.
%    Area (1) is covered by the 
%    MSSM $\tilde{\tau}_R$, (2) by the small impact parameter, (3) by the 
%    kink and large impact parameter searches
%    and (4) by the stable heavy lepton search. 
    The positive-slope hatched area shows the region excluded 
    by the small impact parameter search. The negative-slope 
    hatched area shows the region excluded by the combination
    of the large impact parameter and secondary vertex searches.
} 
  \label{fig:grav:xcl}
\end{figure}

\begin{figure}[htbp]\centering
\epsfxsize=14.5cm
\centerline{\epsffile{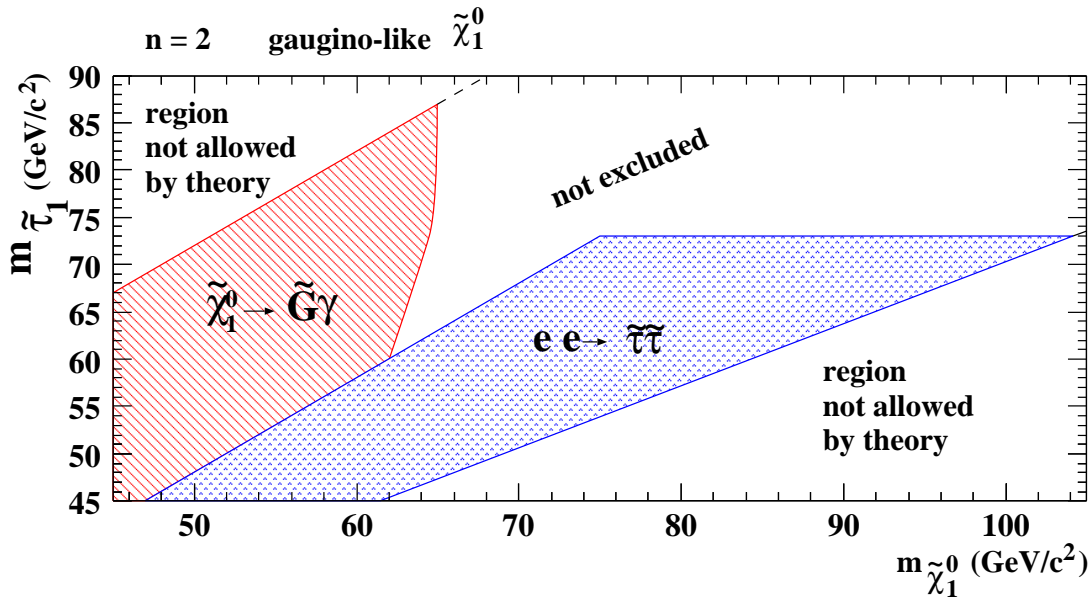}}
\caption[]{Areas excluded at the 95\% C.L. for $n=2$, gaugino-like 
neutralinos and a 40~\eVcc\ gravitino in the 
$m_{\tilde{\chi}_1^0}$\ {\it vs.} $m_{\tilde{\tau}_1}$ plane. 
The  negative-slope dashed
area is excluded by the search for 
$\tilde{\chi}^0_1\rightarrow \gamma \tilde{G}$
 and the point-hatched area by this analysis. 
}
\label{fig:masses_carmen}
\end{figure}

Combining these results with the results of the 
search for MSSM $\tilde{\tau}_R$,
%decays into $\tau$ and neutralino 
allows the exclusion of stau masses below 68.5~\GeVcc\
irrespective of the gravitino mass. The results are shown in
Figure~\ref{fig:grav:xcl}, for $\tilde{G}$ larger than 250~\eVcc\
(not shown in the plot) the limit was 80.0~\GeVcc\, obtained from the
stable heavy lepton search~\cite{Heavyparticles}. 
Following~\cite{Dutta} as in section~\ref{sec:resultados:neutralino}, 
figure~\ref{fig:masses_carmen} 
shows the 95\% C.L. excluded areas for the case of
$n=2$, gaugino-like neutralinos and  $m_{\tilde{G}}=40$~\eVcc\ in the 
$m_{\tilde{\chi}_1^0}$\ {\it vs.} $m_{\tilde{\tau}_1}$ plane. 
The negative-slope dashed area is excluded by
the analysis searching for neutralino pair production followed by the decay
$\tilde{\chi}^0_1\rightarrow \tilde{G}\gamma$. 
The point-hatched
area is excluded by this search taking into account the possibility of
$\tilde{\tau}_L - \tilde{\tau}_R$\ mixing~\cite{bartl}.
The resulting 95\% C.L. lower limit
on the mass of the lightest neutralino is 62~\GeVcc, and that for the 
stau 60~\GeVcc, from the neutralino pair production search. 

By comparing figures~\ref{fig:masses} and~\ref{fig:masses_carmen}
it can be seen that the exclusion power of isolated photon searches 
decreases as the mass of the gravitino increases. It can also be seen that the 
area excluded by the stau pair production searches increases with 
$m_{\tilde{G}}$.

%\newpage

%%%%%%%%%%%%%%%%%%%%%%%%%%%%%%

%%%%%%%%%%%%%%%%%%%%%%%%%%%%%%
%\newpage
\section{Summary}
Lightest neutralino and stau pair production were searched for in the 
context of light gravitino scenarios. It was assumed that the 
$\tilde{\tau}_1$~is
the NLSP and that the $\tilde{\chi}_1^0$ is the NNLSP. 
Both searches were used 
in order to explore the
 ($m_{\tilde{\chi}_1^0},m_{\tilde{\tau}_1}$) plane in different 
domains of the gravitino mass.

The search for neutralino pair production produced two candidate 
events to be compared to
0.77$\pm 0.16$~events expected from the SM background for the samples ranging 
from $\sqrt{s}=161$~GeV up to 183~GeV. An upper limit on the 
corresponding production cross-section between 0.2 and 0.3 pb 
was set at 95\% C.L. in the 
kinematically allowed region. 

The search for the pair production of long lived staus produced 
one candidate for the 
small impact parameter method and none for the large impact 
parameter and vertex methods, whereas totals of
$2.68^{+0.84}_{-0.31}$~and $0.63^{+0.55}_{-0.12}$ events were 
expected  from the simulated SM background, respectively.  
An upper limit on the 
stau pair production cross-section 
was set as a function of its mass and that of the 
gravitino, between 0.1 and 2 pb 
at 95\% C.L. in the 
kinematically allowed region. This result, together with 
the search for staus within the 
MSSM framework and stable stau production, allow the DELPHI collaboration 
to set the lower limit on the mass of 
the $\tilde{\tau}_R$ at 68.5~\GeVcc\ at 95\% C.L..

%         Created on 12-FEB-1998 by dimartino
%-------------------------------------------------------------------
\subsection*{Acknowledgements}
\vskip 3 mm
We would like to thank S.~Nandi, K.~Cheung and C.~Wagner for very useful 
discussions 
in theoretical matters.
 We are greatly indebted to our technical 
collaborators 
%and to the funding agencies for their
%support in building and 
building and  operating the DELPHI detector and to the members
of  the CERN-SL Division for the excellent performance of the LEP collider.\\
We are also grateful to the technical and engineering staffs in our 
laboratories and we acknowledge the support of \\
Austrian Federal Ministry of Science, Research and Arts, \\
FNRS--FWO, Belgium,  \\
FINEP, CNPq, CAPES, FUJB and FAPERJ, Brazil, \\
Czech Ministry of Industry and Trade, GA CR 202/96/0450 and GA AVCR A1010521,\\
Danish Natural Research Council, \\
Commission of the European Communities (DG XII), \\
Direction des Sciences de la Mati$\grave{\mbox{\rm e}}$re, CEA, France, \\
Bundesministerium f$\ddot{\mbox{\rm u}}$r Bildung, Wissenschaft, Forschung 
und Technologie, Germany,\\
General Secretariat for Research and Technology, Greece, \\
National Science Foundation (NWO) and Foundation for Research on Matter (FOM),
The Netherlands, \\
Norwegian Research Council,  \\
State Committee for Scientific Research, Poland, 2P03B00108, 2P03B03311 and
628/E--78--SPUB--P03--023/97, \\
JNICT--Junta Nacional de Investiga\c{c}\~{a}o 
Cient\'{\i}fica 
e Tecnol$\acute{\mbox{\rm o}}$gica, Portugal, \\
Vedecka grantova agentura MS SR, Slovakia, Nr. 95/5195/134, \\
Ministry of Science and Technology of the Republic of Slovenia, \\
CICYT, Spain, AEN96--1661 and AEN96-1681,  \\
The Swedish Natural Science Research Council,      \\
Particle Physics and Astronomy Research Council, UK, \\
Department of Energy, USA, DE--FG02--94ER40817. \\

\newpage
%%%%%%%%%%%%%%%%%%%%%%%%%%%%%%%%%%%%%%%%%%
%% BIBLIO
%%%%%%%%%%%%%%%%%%%%%%%%%%%%%%%%%%%%%%%%%%

\end{document}